\documentclass[fleqn,usenatbib,useAMS]{mnras}

\usepackage{newtxtext,newtxmath}

\usepackage[T1]{fontenc}
\usepackage{ae,aecompl}

\usepackage{graphicx}	
\usepackage{amsmath}	
\usepackage{amssymb}	

\newcommand*{\Rey}{{\rm Re}}
\newcommand*{\Bu}{{\rm Bu}}

\newcommand*{\Ek}{{\rm E}}
\newcommand*{\Ta}{{\rm Ta}}

\newcommand{\Ros}{{\rm Ro}}
\newcommand*{\Pec}{{\rm Pe}}
\newcommand*{\Nus}{{\rm Nu}}
\newcommand*{\Ray}{{\rm Ra}}
\newcommand*{\Rayc}{{\rm Ra_c}}
\newcommand{\Pra}{{\rm Pr}}

\newcommand{\kv}{\ensuremath{\mathbf{k}}}
\newcommand{\rv}{\ensuremath{\mathbf{r}}}
\newcommand{\ve}{{\mathbf{v}}}
\newcommand{\lp}{\ensuremath{\left(}}
\newcommand{\rp}{\ensuremath{\right)}}
\newcommand{\er}{{\bf\hat e_r}}
\newcommand{\et}{{\bf\hat e_\theta}}
\newcommand{\ep}{{\bf\hat e_\varphi}}
\newcommand{\ez}{{\bf\hat e_z}}

\title[Polar dynamics in deep convection models]{Deep model simulation
  of polar vortices in gas giant atmospheres}

\author[F. Garcia et al.]{
Ferran Garcia,$^{1,2}$\thanks{E-mail: f.garcia-gonzalez@hzdr.de}
Frank R. N. Chambers,$^{2}$
and Anna L. Watts$^{2}$
\\
$^{1}$Dpt. of Magnetohydrodynamics, Helmholtz-Zentrum Dresden-Rossendorf, Bautzner Landstra\ss e 400, D-01328 Dresden, Germany\\
$^{2}$Anton Pannekoek Institute for Astronomy, University of Amsterdam, Postbus 94249, 1090 GE Amsterdam, The Netherlands
}

\date{Accepted XXX. Received YYY; in original form ZZZ}

\pubyear{2020}

\begin{document}
\label{firstpage}
\pagerange{\pageref{firstpage}--\pageref{lastpage}}
\maketitle

\begin{abstract}
The Cassini and Juno probes have revealed large coherent cyclonic
vortices in the polar regions of Saturn and Jupiter, a dramatic
contrast from the east-west banded jet structure seen at lower
latitudes.  Debate has centered on whether the jets are shallow, or
extend to greater depths in the planetary envelope.  Recent
experiments and observations have demonstrated the relevance of deep
convection models to a successful explanation of jet structure, and
cyclonic coherent vortices away from the polar regions have been
simulated recently including an additional stratified shallow
layer. Here we present new convective models able to produce
long-lived polar vortices.  Using simulation parameters relevant for
giant planet atmospheres we find flow regimes of geostrophic
turbulence (GT) in agreement with rotating convection theory. The
formation of large scale coherent structures occurs via
three-dimensional upscale energy transfers. Our simulations generate
polar characteristics qualitatively similar to those seen by Juno and
Cassini: they match the structure of cyclonic vortices seen on
Jupiter; or can account for the existence of a strong polar vortex
extending downwards to lower latitudes with a marked spiral
morphology, and the hexagonal pattern seen on Saturn. Our findings
indicate that these vortices can be generated deep in the planetary
interior. A transition differentiating these two polar flows regimes
is described, interpreted in terms of force balances and compared with
shallow atmospheric models characterising polar vortex dynamics in
giant planets. In addition, heat transport properties are
investigated, confirming recent scaling laws obtained with reduced
models of GT.  
\end{abstract}




\section{Introduction}

Zonal (east/west directed) wind circulations are ubiquitous in gas
giant planets. Jupiter and Saturn have strong prograde (eastward)
equatorial jets that extend into the deep interior molecular
envelope~\citep{Kas_et_al18,Gui_et_al18}. In addition, coherent
vortices and related structures are found to be very robust close to
polar latitudes~\citep{Adr_et_al18,Fle_et_al18}. Current modelling of
gas giant flow dynamics involves two different
approaches~\citep{SKAI18}: a shallow-layer scenario in which baroclinic
turbulence is feeding the jets in a thin atmospherical
layer~\citep{LiSc10,MSSFC15,RZS17}; or a deep
scenario~\citep{Bus76,HAW05,HeAu07,HGW15} in which the jets form
quasigeotrophic columns extending down into the molecular
envelope. Which approach is best is still under debate but recent
laboratory experiments~\citep{CAFL17}, reproducing jet properties at
high latitudes in the presence of viscous damping, recent observations
of the deep extension of the Jovian jet streams~\citep{Kas_et_al18},
and analysis of Cassini data from two different pressure levels within
Saturn's polar winds~\citep{Stu_et_al18} all favour the deep approach.

Three-dimensional deep models of the molecular envelope in a rotating
spherical geometry~\citep{HAW05,HeAu07} have provided a strong
background for understanding the mechanisms of jet formation, in which
geostrophic turbulence (GT) is a key issue~\citep{Rhi75}.  The deep
convection models of~\citet{HeAu07} interpreted the transition between
Jupiter and Saturn's flow regimes by considering different spherical
shell widths, pointing out the relevance of the tangent cylinder (a
coaxial cylinder touching the inner boundary at the equator) on the
equatorial jet width. The existence of such an internal boundary for
zonal jets has been one of the main concerns for deep modelling
~\citep{SKAI18}, but recent gravity field measurements by the Juno
mission~\citep{Gui_et_al18} have confirmed that zonal jets extend
deeper into Jupiter's atmosphere. The current explanation for this
internal boundary involves the existence of magnetic field drag acting
on the flow~\citep{LGS08}. Measurements of the electrical conductivity
in the molecular envelope were used to constrain the extent of zonal
jets, giving a lower boundary for Jupiter and Saturn estimated to be
0.96R and 0.86R (R being the planetary radius),
respectively~\citep{LGS08}, which match quite well with those obtained
with recent gravity measurements~\citep{Gui_et_al18}.

Polar coherent vortices in Jupiter were discovered very
recently~\citep{Adr_et_al18} and thus much of the modelling of polar
vortices to date has focused on Saturn. Observations have revealed a
seasonal influence on the morphology of polar flow
structure~\citep{SBDEI17,Fle_et_al18} involving a strong single polar
vortex with associated spiralling arms.  A hexagonal structure
is present near $75^{\circ}$ north latitude at stratospherical levels,
which was believed to be the result of a Rossby wave trapped in the
troposphere~\citep{God88,Fle_et_al18}.  Existing models of Saturn's
hexagonal pattern and polar vortices assume shallow jet
structures~\citep{MSSFC15,RZS17} and can explain key features of the
hexagonal pattern such as its latitudinal location, almost steady
azimuthal drift and the absence of large vortices close to the structure.
However, there is debate about whether this hexagonal jet has a deep
origin, as the long term observations from the Cassini mission
suggest~\citep{Bai_et_al09,San_et_al14}. In the case of Jupiter, the first
close-up images of north pole revealed the coexistence of cyclonic
vortices, oval-shaped structures, filamentary regions and thin linear
features elongated more than $30^{\circ}$ in the azimuthal direction,
concentrated from $60^{\circ}$ latitude
polewards~\citep{Ort_et_al17}.  Subsequent Juno data analysis confirmed
the persistence of these flow structures: 8 vortices surround the main
cyclonic north polar vortex, whereas only 5 vortices are found
surrounding the main cyclonic south polar vortex~\citep{Adr_et_al18}.

Direct numerical simulations (DNS) of shallow atmospheric
models~\citep{Sco11,OEF15,OEF16} have been used to understand polar
atmospheric dynamics of giant planets (see the comprehensive review
of~\citealt{Say_et_al18} in the case of Saturn), especially the case
of polar cyclones. The latter studies consider a source term which
accounts for small-scale features produced by moist convective
storms~\citep{OEF15,OEF16} and drives turbulence. The idea behind this
approach is that cyclonic vortices form as a result of poleward flux
of cyclonic vorticity caused by the beta-drift, i.e. the advection of
the background potential vorticity by the storm circulation. The moist
baroclinic forcing is then responsible for the generation of a
barotropic vortex aligned in the vertical
direction~\citep{OEF15}. This study first demonstrated that the vortex
features of the flow, i.~e. the appearance (or not) of a single vortex
centered at the poles, is controlled by the ratio of two
characteristic sizes, namely, that of the small vortices and the
planetary radius. The recent study by \citet{BSD19} provides further
support for the mechanism which distinguishes the specific
characteristics of polar dynamics occurring in Jupiter (a ring of
cyclones surrounding the poles) and Saturn (single and strong cyclone)
studied in~\citet{OEF15,OEF16}. The key parameter which controls the
transition between the latter scenarios is the Burger number, defined
as the ratio between the Rossby deformation radius and the planetary
radius~\citep{BSD19}. As the Burger number is decreased a transition
from Saturn-like to Jupiter-like polar cyclones is observed.

Rotating Rayleigh-B\'enard convection between plane boundaries
constitutes a reasonable approximation for studying polar dynamics in
gas giant atmospheres as the convective region is very thin when
compared to the planetary radius. Within this canonical framework much
theoretical and numerical work has recently been
done by~\citet{JRGK12,Jul_etal12,Rub_etal14} to investigate the regime of
GT in terms of a 3D reduced model including boundary effects. The
latter studies identified the physical mechanisms involved in the
formation of large-scale coherent structures surrounding the rotation
axis as a result of a fully three-dimensional convective
forcing~\citep{Rub_etal14}. The associated efficient mixing within the
fluid reduces the heat transport and determines its scaling
law~\citep{Jul_etal12}. According to~\citet{JRGK12} the realisation of
the GT regime, in which the flow is weakly dependent on the axial
coordinate, depends strongly on the input parameters, particularly the
Prandtl number associated with the thermal properties of the
fluid. For smaller Prandtl numbers the GT regime is favoured, in
agreement with our results.

The formation of these coherent structures at polar latitudes in giant
planets remains to be reproduced within the context of deep convection
modelling. This is the focus of the present study, based on three
dimensional DNS of Boussinesq thermal convection in rotating spherical
shells without any symmetry assumption. In Sec.~\ref{sec:mod} the
model equations, numerical method and the physical parameters of the
models are described. Sec.~\ref{sec:tran} constitutes the main part of
the study. It includes the description of the flow patterns and their
time averaged properties, the investigation of a condensate process
and the transition between polar regimes, the study of the force
balance and the thermal properties, and the characterisation of a
Saturn-like hexagonal pattern. Finally, a summary of the main results
and future investigations is provided in Sec.~\ref{sec:con}. In this
closing section, we also include a discussion of the validity of the
approximations and approaches taken, when applying our findings to
the giant planets.

\section{The model}
\label{sec:mod}

\subsection{Governing equations and numerical method}
\label{sec:eq}

A homogeneous fluid with density $\rho$ and constant physical
properties --thermal diffusivity $\kappa$, thermal expansion
coefficient $\alpha$, and dynamic viscosity $\mu$-- is considered, to
study thermal convection in a rotating spherical shell. The latter is
defined by the inner and outer radius $r_i$ and $r_o$, and a
constant angular velocity ${\bf \Omega}=\Omega {\kv}$ about the
vertical axis. Gravity acts in the radial direction ${\bf g}=-\gamma
\rv$ ($\gamma$ is constant and $\rv$ the position vector) and the
relation $\rho=\rho_0(1-\alpha(T-T_0))$ is assumed (the Boussinesq
approximation), but just in the gravitational term. For the other
terms appearing in the equations a reference state $(\rho_0,T_0)$ is
considered (see for instance~\citealt{Ped79,Cha81}).

At the boundaries, which are perfectly conducting, a temperature
difference is imposed $\Delta T=T_i-T_o$, $T(r_i)=T_i$ and
$T(r_o)=T_o$. Stress-free boundary conditions for the velocity field
are appropriate for planetary atmospheres~\citep{Chr02,HAW05}. The
mass, momentum and energy equations are written in the rotating frame
of reference as in~\citet{SiBu03} and expressed in terms of velocity
($\ve$) and temperature ($\Theta=T-T_c$) perturbations of the basic
conductive state $\ve=0$ and $T_c(r)=T_0+\eta d \Delta
T(1-\eta)^{-2}r^{-1}$, $\eta=r_i/r_o$ being the aspect ratio,
$d=r_o-r_i$ being the gap width, and $T_0=T_i-\Delta T(1-\eta)^{-1}$
being a reference temperature.  With units $d$ for the distance,
$\nu^2/\gamma\alpha d^4$ for the temperature, and $d^2/\nu$ for the
time, the governing equations read
\begin{align}
&\nabla\cdot\ve=0,\label{eq:cont}\\
&\partial_t\ve+\ve\cdot\nabla\ve+2\Ta^{1/2}\kv\times\ve = 
-\nabla p^*+\nabla^2\ve+\Theta\rv,\label{eq:mom}\\
&\Pr\lp\partial_t \Theta+\ve\cdot\nabla \Theta\rp= \nabla^2
\Theta+\Ray \eta (1-\eta)^{-2}r^{-3} \rv\cdot\ve,  \label{eq:ener}
\end{align}
being $p^*$ is a dimensionless scalar containing all the potential
forces. Centrifugal effects are neglected by assuming $\Omega^2/\gamma
\ll 1$. This is usual for the case of spherical shell
convection~\citep{ScZh00}. Four non-dimensional parameters -the aspect
ratio $\eta$, the Rayleigh $\Ray$, Prandtl $\Pr$, and Taylor $\Ta$
numbers- summarise the physics of the problem. The Prandtl number
characterises the relative importance of viscous (momentum)
diffusivity to thermal diffusivity, the Taylor number is the ratio
between rotational and viscous forces, and the Rayleigh number $\Ray$
is associated with buoyancy forces and thermal forcing. The parameters
are defined by
\begin{equation}
  \eta=\frac{r_i}{r_o},\quad
  \Ray=\frac{\gamma\alpha\Delta T d^4}{\kappa\nu},\quad
  \Ta^{1/2}=\frac{\Omega d^2}{\nu},\quad
  \Pr=\frac{\nu}{\kappa}.
\label{eq:param}
\end{equation}

To solve the model equations~\ref{eq:cont}-\ref{eq:ener} with the
prescribed boundary conditions a pseudo-spectral method~\citep{Til99}
is used (see~\citealt{GNGS10} for a detailed description). In the
radial direction a collocation method on a Gauss--Lobatto
mesh~\citep{SGN16} is considered and spherical harmonics are employed
in the angular coordinates. In the Boussinesq approximation the
toroidal/poloidal decomposition for the velocity field can be
used~\citep{Cha81}. The code is parallelized in the spectral and in
the physical space using OpenMP directives. Optimized libraries
\citep[FFTW3,][]{FrJo05} for the fast Fourier transforms FFTs in
longitude and matrix-matrix products \citep[dgemm GOTO,][]{GoGe08} for
the Legendre transforms in latitude are implemented for the
computation of the nonlinear terms. Aliasing is properly removed in
the pseudo-spectral transform method~\citep{Ors70}.

High order implicit-explicit backward differentiation formulas
\citep[IMEX--BDF,][]{GNGS10} are used for the time integration of the
discretized equations. The nonlinear terms are considered explicitly, to
avoid solving nonlinear equations at each time step. However, the
Coriolis term is treated fully implicitly and this allows a time
integration with larger time steps~\citep{GNGS10}.

\subsection{Model parameters}
\label{sec:par}

\begin{table}
  \begin{center}
    \begin{tabular}{lccc}
\vspace{0.1cm}           
Model              & 1                & 2\\
\hline\\[-10.pt]
$\Pra$             &$0.01$             &$0.01$           \\       
$\Ta$              &$10^{11}$           &$10^{11}$         \\   
$\Ray$             &$5\times 10^5$      &$8\times 10^5$   \\    
$\Ray/\Rayc$       &$10.8$              &$17.2$           \\    
$\Ray^*$           &$5\times 10^{-3}$    &$8\times 10^{-3}$ \\   
$\overline{\Rey}$  &$4.3\times 10^3$    &$6.3\times 10^3$  \\   
$\overline{\Ros}$  &$1.4\times 10^{-2}$ &$2.0\times 10^{-2}$\\  
$\overline{\Pec}$  &$4.3\times 10^{1}$  &$6.3\times 10^{1}$ \\  
\hline
\end{tabular}
    \caption{Input and output parameters for the different models: the
      Prandtl $\Pra$, Taylor $\Ta$, Rayleigh $\Ray$ and modified
      Rayleigh $\Ray^*=\Ray\Ta^{-1}\Pra^{-1}$ numbers. The Reynolds
      $\overline{\Rey}=\overline{\sqrt{2K}}$, Rossby
      $\overline{\Ros}=\overline{\Rey}/\Ta^{1/2}$, and P\'eclet
      $\overline{\Pec}=\overline{\Rey}\Pra$ numbers are based on the
      volume and time averaged kinetic energy $\overline{K}$. The time
      averages cover 1500 and 1300 planetary rotations for Model 1 and
      2, respectively.}
  \label{tab_param}
  \end{center}
\end{table}


\begin{figure*}
  \includegraphics[width=0.95\textwidth]{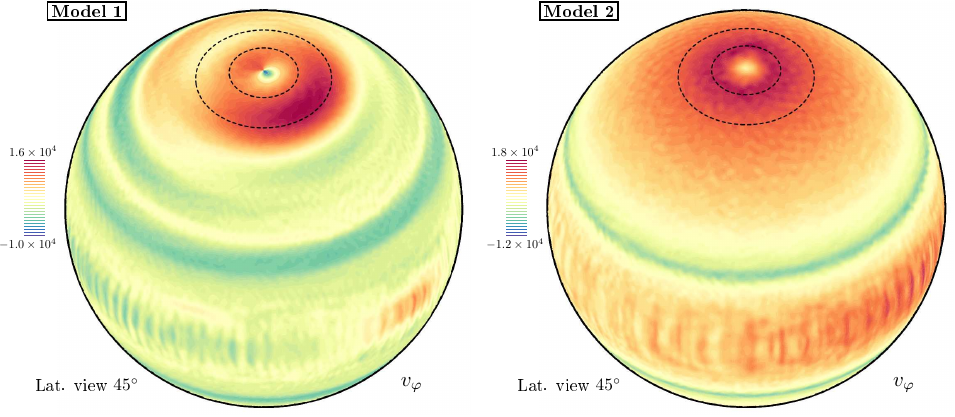}\\[2.mm]
  \includegraphics[width=0.95\textwidth]{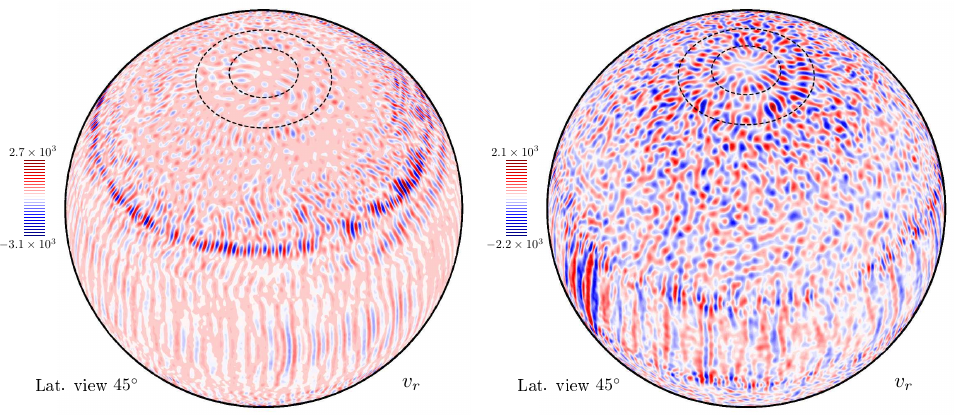}
\caption{{\bf{Velocity patterns of numerical models.}}  Snapshots of
  azimuthal velocity $v_{\varphi}$ at the outer surface $r=r_o$ (1st
  row), and of radial velocity at $r=r_i+0.5d$ (2nd row), both viewed
  from a latitude of $45^{\circ}$. Left column is for Model 1 and
  right column is for Model 2. Positive (negative) values are marked
  with red (blue). Parallel circles at latitudes
  $-80^{\circ},-70^{\circ}$ and $70^{\circ},80^{\circ}$ are marked
  with dashed lines.}
\label{fig_sph_vel}
\end{figure*}

Two different models of Boussinesq thermal convection are considered
to explore the transition from Saturn-like (Model 1) to Jupiter-like
(Model 2) polar dynamics. In these two models the parameters $\eta$,
$\Ta$ and $\Pr$ are the same. They are similar to previous numerical
models of gas giant atmospheres~\citep{HAW05,HeAu07,HGW15} focusing on
the understanding of jet dynamics away from the polar regions.  We
select an aspect ratio $\eta=r_i/r_o=0.9$ which roughly falls between
the values estimated for Jupiter and Saturn~\citep{LGS08}. We set a
sufficiently large Taylor number, $\Ta=10^{11}$ - as used
in~\citet{HAW05,HeAu07,HGW15} - and a Prandtl number $\Pra=10^{-2}$,
which is a reasonable regime for gas giant
atmospheres~\citep{Fre_et_al12} due the uncertainty in determining
$\Pr$. A key difference between the current polar models and those
of~\citet{HAW05,HeAu07} and \citet{HGW15} devoted to lower latitudes,
is the selection of a small Prandtl number, which strongly influences
the type of convection~\citep{SiBu03,KSVC17,GCW19}. In our case
convection directly onsets at high latitudes~\citep{GCW18} whereas
previous deep models require a strong nonlinear regime to develop
polar convection.

With an aspect ratio of $\eta=0.9$, $\Ta=10^{11}$ and $\Pra=10^{-2}$
convective onset takes the form of nonaxisymmetric polar modes
confined to large latitudes~\citep{GSN08,GCW18} at a critical Rayleigh
number $\Rayc=4.65\times 10^4$. The main difference between our two
models comes from the Rayleigh number. Model 1 (M1) corresponds to
$\Ray=5\times 10^5$ ($\Ray/\Rayc=10.8$), whereas Model 2 (M2)
corresponds to a slightly larger $\Ray=8\times 10^5$
($\Ray/\Rayc=17.2$). For low Prandtl and large Taylor numbers even a
small degree of supercriticallity provides strong turbulent
flows~\citep{KSVC17}, characterised by a predominance of advection,
rather than diffusion, in the temperature transport.

The DNS presented here have spatial resolutions very similar to
those used in previous deep convection models for the same range of
parameters (Eg.~\citealt{HAW05,HGW15}). Specifically, we consider spherical
harmonics up to order and degree $L_{\max}=500$ and a radial mesh with
$n_r=60$ points. In contrast to previous studies, however, our
numerical simulations do not rely on the use of azimuthal symmetry
constraints, nor on imposed hyperviscosity, allowing us to capture low
wave number coherent dynamics in the flow. These are otherwise
filtered if $m=m_d>6$-fold symmetry constraints are imposed. However,
the system of equations has in return dimension
$n=(3L_{\text{max}}^2+6L_{\text{max}}+1)(N_r-1)\approx 4.4\times
10^7$, making the numerical integration of Models 1 and 2 very
challenging.

Both models are evolved for around one viscous time unit ($1.5$ for M1
and $1.3$ for M2), including a large initial transient, with a time
step of $\Delta t=2\times 10^{-7}$.  The initial condition for
integrating Model 1 corresponds to a transient flow at the same
$\Pr=10^{-2}$ and $\Ta=10^{11}$ but with $\Ray=2.32\times 10^{5}$. The
latter simulation has been initialised from a model at $\Pra=3\times
10^{-3}$, $\Ta=10^7$, and $\Ray=2\times 10^3$ ~($\Ray/\Rayc=6.6$)
obtained from the chaotic polar flows described in~\citet{GCW19}.
Model M2, with larger $\Ray$, is obtained from M1. This is a common
strategy in deep convection models~\citep{Chr02,HAW05,GSN14}.  Because
of the large dimension of the system (which has been integrated
without any symmetry assumption) and the small time step required to
obtain the numerical simulations, only two simulations are
presented. These are however some of the highest resolution convective
models, in rotating thin shells, performed to date.

The two models exhibit features of turbulent flows that occur in gas
giant atmospheres. The values shown in Table~\ref{tab_param} of the
time and volume averaged Rossby $\overline{\Ros}$ (measuring the
balance between Coriolis and inertial forces) or Reynolds
$\overline{\Rey}$ (a measure of turbulent flows) numbers are in
agreement with those given in previous studies on gas giant
atmospheres, for instance those based on the peak zonal flow
velocities, given in Table II of~\citet{HeAu07}. The modified Rayleigh
number $\Ray^*=\Ray\Ta^{-1}\Pra^{-1}\ge 0.005$, measuring the ratio of
buoyancy to Coriolis forces, matches reasonably as well. In addition,
the P\'eclet number (measuring the ratio of temperature transport by
advection and diffusion) is larger than $10$ (see~\citealt{KSVC17})
a value which marked the transition between laminar and turbulent
flows at low $\Pra$ and large $\Ta$.

\section{Deep polar flow dynamics}
\label{sec:tran}

In this section we explore and analyse in detail the new regime of
polar convection exhibited by our two models, M1 and M2, focusing on
its relation to the polar dynamics observed on Saturn and
Jupiter. First, we provide a description of the main characteristics
of the flow patterns, followed by an analysis of time averaged flow
latitude profiles. Second, we identify a turbulent energy transfer
mechanism by means of kinetic energy spectra, which in turn are
compared with results from the Cassini mission. Third, the appearance
of different coherent polar structures is interpreted within recent
theoretical results of 3D rotating convection, and related to shallow
water models of Saturn and Jupiter. We then analyse the force balance
and the properties of thermal transport.  This section closes with the
description of a Saturn-like hexagonal flow pattern, surrounding the
north pole, observed in M1.

\subsection{Description of flow patterns}
\label{sec:fl_pat}

\begin{figure}
\includegraphics[width=0.49\textwidth]{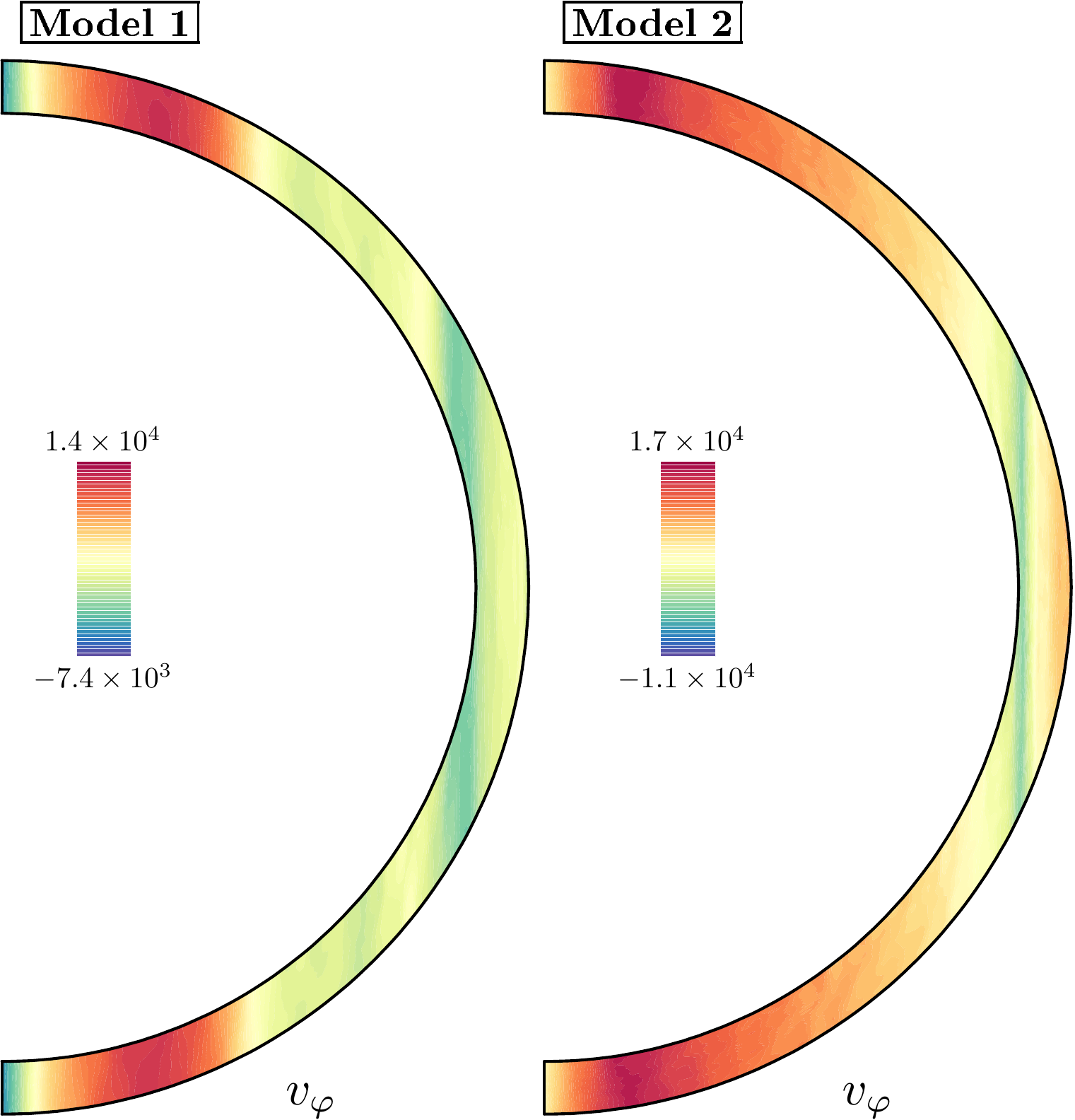}\\
\includegraphics[width=0.49\textwidth]{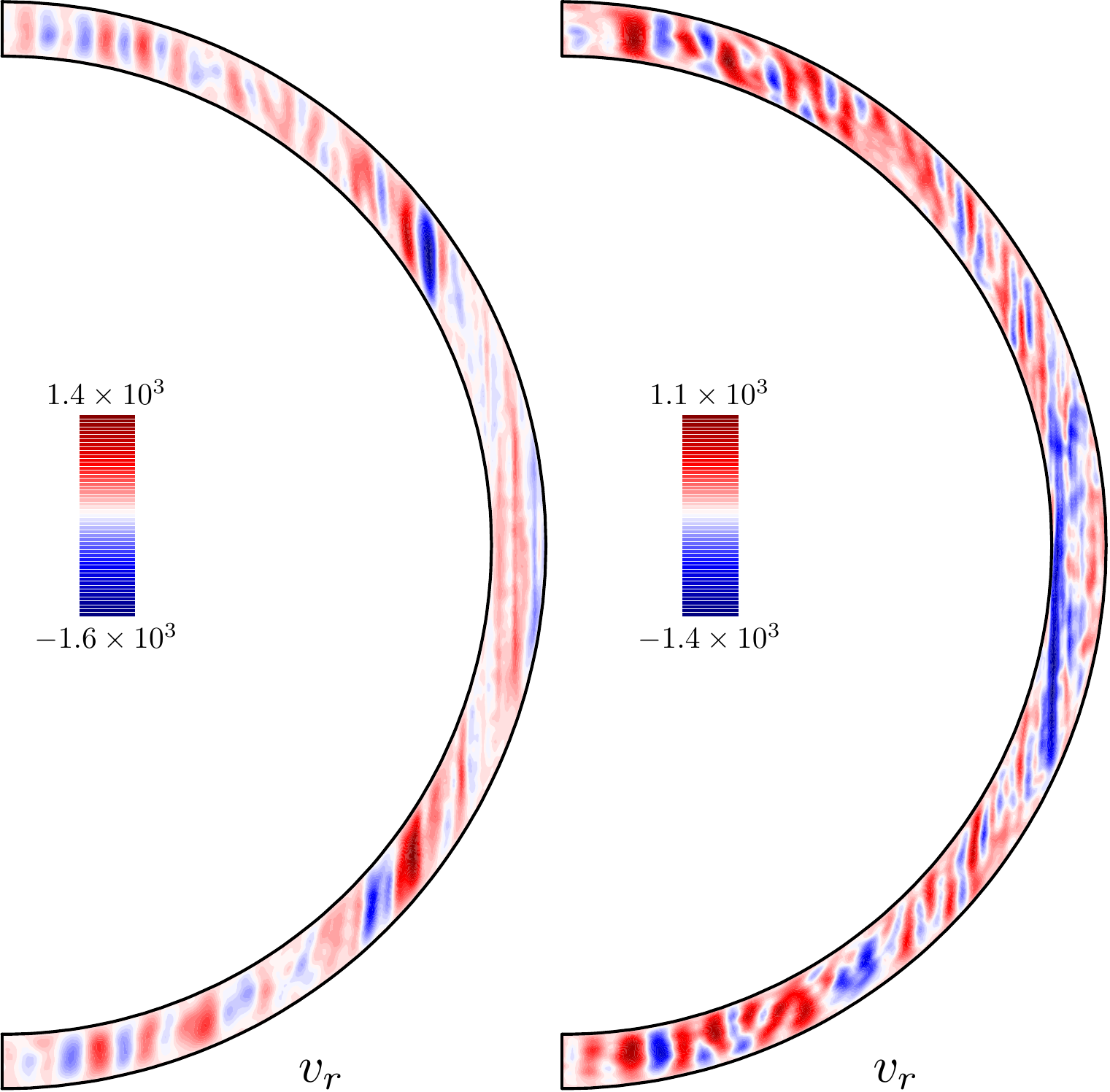}  
\caption{{\bf{Velocity patterns of numerical models.}}  Snapshots of
  azimuthal velocity $v_{\varphi}$ (1st row) and of radial velocity
  (2nd row) on a meridional section for  Model 1 (left column) and
  Model 2 (right column). Positive (negative) values are marked with red
  (blue).}
\label{fig_ver_vel}
\end{figure}

\begin{figure*}
\includegraphics[width=0.9\textwidth]{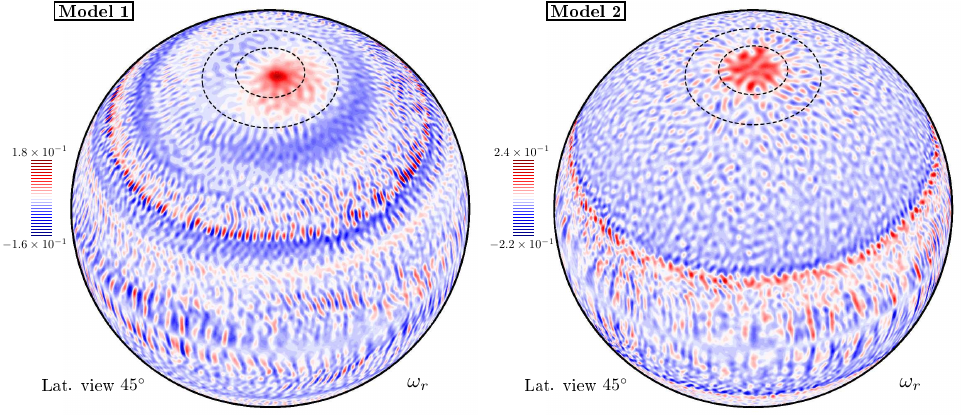}\\[2.mm]
\includegraphics[width=0.9\textwidth]{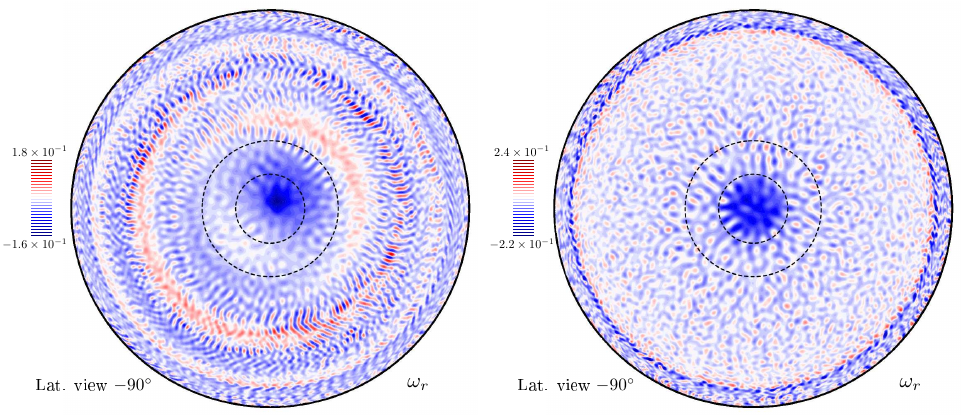}  
\caption{{\bf{Coherent polar vortices of numerical models.}}
  Snapshots of radial vorticity $\omega_r$ (in planetary rotation
  units) at the outer surface $r=r_o$ viewed from a latitude of
  $45^{\circ}$ (left column) and from a latitude of $-90^{\circ}$
  (right column). The 1st row corresponds to Model 1 and the 2nd row
  to Model 2. Cyclonic (anticyclonic) radial vorticity is red (blue)
  in the northern hemisphere and blue (red) in the southern
  hemisphere. Parallel circles at latitudes $-80^{\circ},-70^{\circ}$
  and $70^{\circ},80^{\circ}$ are marked with dashed lines.}
\label{fig_sph_vor}
\end{figure*}

Figure~\ref{fig_sph_vel} displays snapshots of the contour plots of
the azimuthal velocity $v_{\varphi}$ at the outer sphere and the
radial velocity $v_r$ at the sphere with radius $r=(r_i+r_o)/2$
(i.e. in the middle of the shell) for both models. The spheres are
viewed from $45^{\circ}$ latitude, and the $70^{\circ}$ and
$80^{\circ}$ parallels surrounding the north pole are marked with a
dashed line. The flow is strongly axisymmetric for both models (see
first row of Fig.~\ref{fig_sph_vel}) as is typical in stress-free
convection models~\citep{AuOl01,Chr02}, with a noticeable wide positive
(prograde) equatorial band which is more evident for M2. This is
typical for flows with $\Ray^* \ll 1$ (Eg.~\citealt{AHW07}) in which the
Coriolis forces dominate over buoyancy effects. In this case Reynolds
stresses associated with columnar convection are feeding the zonal flow
at mid and low latitudes~\citep{Chr02}. In contrast to the latter
studies, strong prograde zonal flows are produced at high latitudes in
models M1 and M2.

The main difference between the models is in the latitudinal extent
of these high latitude zonal circulations and their topology, which for
M1 has a noticeable $m=1$ azimuthal wave number component. The
existence of small scale vortices associated with turbulent flows is
best seen in the contour plots of the radial velocity (second row of
Fig.~\ref{fig_sph_vel}). In this case very small scale flow structures
at high latitudes coexist with large scale cells elongated in the
axial direction near the equator. As evidenced by the figure, the
radial vortices within the $70^{\circ}$ parallel are stronger in the
case of M2. The geostrophic character of the flows, i.e. the tendency
of the flow to align with the rotation axis, is reflected in
Figure~\ref{fig_ver_vel} on the meridional sections of $v_{\varphi}$
and $v_r$. The zonal circulation of polar jets extends down to the inner
boundary and is almost independent of the $z$ coordinate. This is
 fulfilled for the small as well as the large scale radial
vortices. Notice how these large scale cells are only allowed to grow
outside the tangent cylinder.

\begin{figure*}
  \includegraphics[scale=2.]{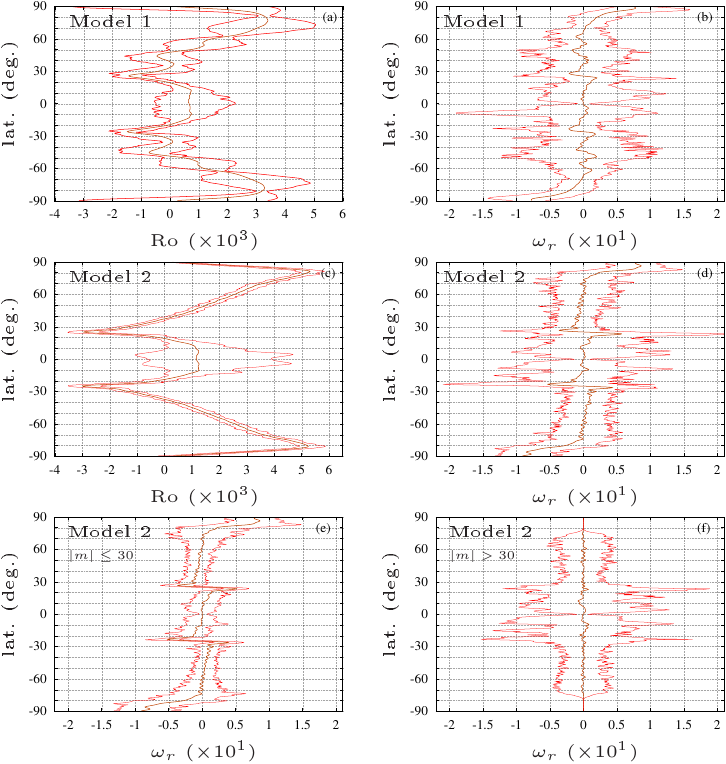}
\caption{{\bf{Flow latitude profiles.}}  (a,c) Time averages of the
  Rossby number $\Ros= v_{\varphi}/\Omega r_o$ (solid line)
  ($v_{\varphi}$ is the azimuthal velocity) versus latitude in
  degrees. The thin dashed lines mark the maximum and minimum values
  of $\Ros$ at each latitude.  (b,d) Time averages (solid line) and
  minimum/maximum values (thin dashed line) of the radial vorticity
  $w_r$ in planetary rotation units. (e,f) display the radial
  vorticity for Model 2 when considering the $|m|\leq 30$ or $|m|>30$
  components of the flow.}
\label{fig_lat_prof}   
\end{figure*}

Figure~\ref{fig_sph_vor} shows the radial vorticity patterns on the
outer sphere for M1 and M2. This figure displays a transition, in
agreement with shallow layer models~\citet{OEF15,OEF16,BSD19}, in
which a single polar vortex (M1) is divided into smaller vortices (M2)
as the thermal forcing ($\Ray$) is increased.  The flow structure of
M1, seen in the first row of Fig.~\ref{fig_sph_vor}, reveals key
features similar to those that appear in Saturn's
atmosphere~\citep[Cassini mission,][]{SBDEI17}. There is a single
strong cyclonic vortex almost centered at both poles and with
spiralling morphology. The polar vortices are surrounded by circular
belts containing small vortices far away from $\pm 70^{\circ}$
latitude. These two strong cyclones are present throughout the
simulation (at least one viscous time unit) in agreement with the
observations~\citep{Fle_et_al15}.

In contrast to M1, the DNS corresponding to M2 (2nd row of
Fig.~\ref{fig_sph_vor}) has long-lived polar structures resembling the
polar features observed in Jupiter's
atmosphere~\citep{Adr_et_al18}. It reveals the prevalence of several
strong cyclonic coherent vortices close to the north as well as south
pole, some of them with an oval shape. Azimuthally elongated
structures exist, in agreement with the observations of the Juno
mission~\citep{Ort_et_al17,Adr_et_al18}. In addition, far away from
the polar regions, circular vortices are arranged in parallel lines
(east-west) and the strongest lie very close to $25.8^{\circ}$
latitude, in either the northern or southern hemisphere as previously
simulated in~\citet{HGW15} with stratified convection models.

We note that Figure~\ref{fig_sph_vor} evidences a multimodal structure
as found in the rotating convection experiments of~\citet{ABGHV18} at
similar $\Pr$ number. Modes with $|m|\leq 30$ are responsible for long
lived polar vortices, which extend down to the inner boundary, as well
as low latitude large-scale cells (meridional sections of
Fig.~\ref{fig_ver_vel}). In contrast, high wave numbers $|m|>30$ only
contribute far away from the latitude circles $\pm 70^{\circ}$. For
this component of the flow two different sizes of vortex are observed:
those elongated in the latitudinal direction near equatorial latitudes
and those with oval shape appearing around $\pm 24^{\circ}$ latitudes.

\subsection{Time averaged latitude profiles}
\label{sec:ta_lat}


The latitude profiles at the outer surface shown in
Fig.~\ref{fig_lat_prof} mark the latitude position of the key features
for both models seen in the previous figures. The time averaged Rossby
number $\Ros= v_{\varphi}/\Omega r_o$ (ratio of inertial to Coriolis
forces) and radial vorticity $w_r$ are plotted versus latitude for
both models. The time series span a interval of around 1300 planetary
rotations for Model 1 and 1500 for Model 2, with a time step of around
20 planetary rotations in both models. The time averaged zonal Rossby
number $\Ros_z=\langle v_{\varphi} \rangle/\Omega r_o$ (where $\langle
v_{\varphi} \rangle$ is the azimuthal average of the azimuthal
velocity) is not displayed since it is almost equal to the time
averaged $\Ros$, i.e.  nonaxisymmetric $v_{\varphi}$ time fluctuations
tend to cancel out. For both models the Rossby number $\Ros$ is
clearly large close to the poles (see Fig.~\ref{fig_lat_prof}, panels
(a,c)) and remains smaller at equatorial latitudes. This means that
our modelling is not reasonable for the study of dynamical behaviour
of gas giant planets at low latitudes~\citep{HeAu07,HAW05,HGW15},
although it reproduces qualitatively several key features of the
equatorial wind such as its latitudinal extent, the negative peaks
close to $\pm 25^{\circ}$ latitudes and even a noticeable decrease of
magnitude near the equator (see Fig.~\ref{fig_lat_prof}(c)). The polar
dynamics exhibited by the present models is a new feature not
previously observed in rotating thermal convection in spherical
shells. There are two peaks of $\Ros$ around $76^{\circ}$ and
$-78^{\circ}$ for Model 1 and they are located closer to the poles
$81^{\circ}$ and $-81^{\circ}$ for Model 2. Indeed, the time
dependence away from $\pm 20^{\circ}$ latitudes is significantly more
vigorous for Model 1. For this model the maximum and minimum curves of
$\Ros$ clearly display two latitudes $\pm 63^{\circ}$ and $\pm
81^{\circ}$ for which $\Ros$ remains roughly constant. Besides a
transition between different polar vortex characteristics between
Model 1 and 2, there is also a transition between equatorial
behaviour: the time dependence of $\Ros$ for Model 2 is enhanced near
the equator ($\pm 20^{\circ}$ latitudes) providing maximum values of
$\Ros$ around $\pm 5^{\circ}$ latitude, comparable to the peaks near
the poles (see Fig.~\ref{fig_lat_prof}(c)).

The radial vorticity latitude profiles shown in Fig.~\ref{fig_lat_prof}(b,d)
clearly display a strong time dependence of vorticity. The main
characteristic featured by both models is the strong cyclonic
vorticity near both poles. For Model 1 the polar jet extends down to
$\pm 70^{\circ}$ latitudes (where $w_r$ becomes zero) and a similar
extent, down to $\pm 73^{\circ}$, is reached for Model 2. For the
latter, the relative maximum/minimum near $\pm 96^{\circ}$ marks the
position of the several circumpolar vortices. The radial vorticity of
Model 2 exhibits stronger oscillations at lower latitudes, especially
around $\pm 24^{\circ}$, where the time average amplitude is
noticeable.  The relative extrema at intermediate latitudes
$\theta\in[25^\circ,70^\circ]$ are larger for Model 1, indicating that
vortex activity is more vigorous (compare also the contour plots seen in
Fig.~\ref{fig_sph_vor}). The bottom row of Fig.~\ref{fig_lat_prof} evidences the
different mode contribution to vortex activity in Model 2 by
considering only the spherical harmonics with $|m|\le 30$ or with
$|m|> 30$. The time averaged $w_r$ of the latter component is nearly zero,
meaning that the main contribution to mean vorticity comes from low
order modes. Indeed, the time dependent $w_r$ is zero near the poles
for the $|m|> 30$ component of the flow, and thus the latter does not
contribute to the origin of the polar dynamics.

Quantitatively, our models underestimate the zonal wind amplitudes
measured in gas giant planets. For instance, in the case of
Saturn~\citep{Say_et_al18} the peak zonal winds close to the poles
(around $\pm85^\circ$ latitude,~\citealt{SHPR06}) are around $150$ m/s
giving rise to $\Ros\sim 0.01$ which is roughly three times that
measured for Model 1. As commented previously, values for the
parameters are not so far from those obtained from the DNS
of~\citet{HeAu07,HAW05,HGW15} (at larger $\Pr$) modelling wind jets at
moderate and low latitudes. However, our models exhibit larger $\Ros$
and circular vortices at high latitudes, in contrast to previous
modelling. In addition, preliminary results for models with larger
$\Ray$ point to a decrease of zonal flows and a disappearance of
coherent structures in the polar regions. Thus there exists a flow
transition, in which zonal flow and vortices at high latitudes
progressively weaken, in a parameter regime relevant for giant
planetary atmospheres. This transition also occurs at lower rotation
rates when polar modes are preferred at the onset~\citep{GCW19}. 
Understanding this transition may provide a simple explanation for
the existence of polar convective coherent vortices in planetary
atmospheres.

\subsection{Kinetic energy spectra and energy transfer mechanism}
\label{sec:en_spec}

\begin{figure}
\includegraphics[width=0.45\textwidth]{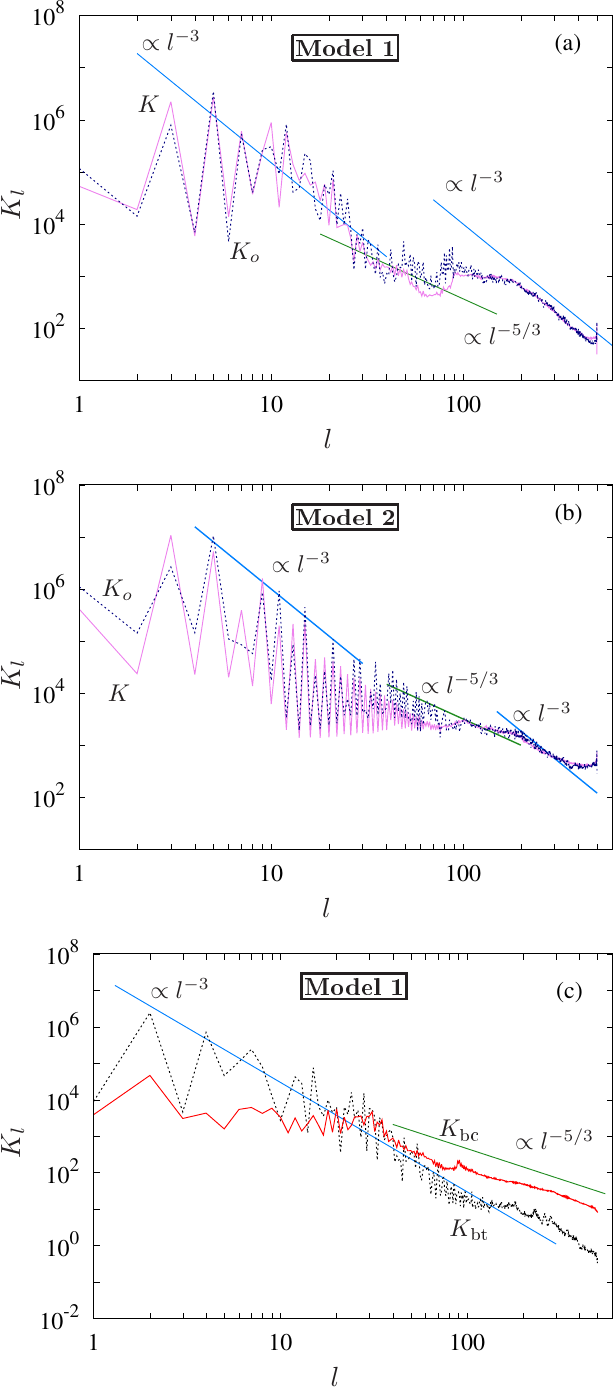}
\caption{{\bf Energy spectra of numerical models.}  (a) Kinetic
  energy spectra versus the spherical harmonic degree $l$ for Model
  1. The volume averaged energy (dashed line) and the energy at the
  outer surface (solid line) are considered. (b) Same as (a) but for
  Model 2. (c) Kinetic energy spectra of the barotropic (dashed line)
  and baroclinic (solid line) flow versus the spherical harmonic
  degree $l$ for Model 1.}
\label{fig_ener_spec}
\end{figure}

An analysis of the kinetic energy distribution among the spherical
harmonic modes with different degree $l$ is performed, to elucidate the
inverse cascade mechanism for these three-dimensional rapidly rotating
flows~\citep{Rub_etal14}. In addition, kinetic energy spectra are a
tool which helps to validate the spherical harmonic resolution of a
given DNS. A rule of thumb~\citep{COG99,KuCh02} indicates that a model
is well resolved if the kinetic energy decreases by more than two
orders of magnitude from the maximum to the smallest wave-length
scale.

Figure~\ref{fig_ener_spec}(a,b) displays the volume averaged kinetic
energy $K$ (dashed line) and the surface averaged kinetic energy $K_o$
at $r=r_o$ (solid line) of a particular mode with degree $l$ for both
models M1 and M2. Notice that the energies of few harmonics just
before the maximum spectral degree $L_{\text{max}}=500$ increase a
little bit. This behaviour is present in some three-dimensional DNS in
rotating spherical geometry, but it has been shown (see~\citealt{COG99}
for the geodynamo case) that a reasonable energy increase at the tail
does not significantly alter the large-scale structure of the flow.

The spectra exhibit slopes of $-5/3$ and $-3$ which are in reasonable
agreement with those obtained from 2D maps of velocity measurements
extracted from Cassini images~\citep{ChSh11}. Furthermore, the
distribution of energy among the different degree $l$ displayed in
Fig.~\ref{fig_ener_spec}(a,c) resembles the distribution exhibited
on Fig.5(a) of~\citet{ChSh11}, constructed from Jovian velocity
measurements.  Because the flow has a strong equatorially symmetric
component, odd $l$ have larger energy than even $l$, especially for low
values, giving rise to the spiking behaviour of the spectra. Note
that~\citet{ChSh11} computes the power spectra of kinetic energy,
rather than the energy content of each mode, so that even $l$
have larger energy. The length scale marking the change of slope from
$-5/3$ to $-3$ of the Jovian spectra of~\citet{ChSh11}, assumed to be
the scale at which energy is injected to the system, is in the
interval $70-140$ depending whether global or eddy kinetic energy are
considered. In our case the situation is similar and the $-5/3$ slope
is lost from $l\approx 60$ in the case of M1 or from $l\approx 120$ in
the case of M2.

In the context of three-dimensional Rayleigh-B\'enard rotating
convection an upscale energy transfer mechanism, known as the
condensation process, has recently been identified in the regime of
GT~\citep{Rub_etal14}. Large scale barotropic vortices, fulfilling the
Taylor-Proudman constraint, are generated through an inverse cascade
mechanism. Small scale convective motions act as an energy source for
these barotropic vortices, which in turn organise these small
convective structures. This is reflected in the energy spectra of the
flow when considering the barotropic and baroclinic components
separately. Flow motions in regions close to the poles in the case of
three dimensional rotating spherical shells can be approximated to
flow motions in a three-dimensional rotating plane layer if the shell
is sufficiently thin. Because this is the case of M1 and M2, we may
compare our kinetic spectra with those obtained on a plane
geometry. Following~\citet{Rub_etal14} we compute the power spectral
energy of the barotropic and baroclinic components of the flow. Close
to the poles we may assume $\er\approx\ez$ (the radial direction is
almost parallel to the axis of rotation) and the plane spanned by
$\et$ and $\ep$ is parallel to the $x-y$ plane. We define the
barotropic kinetic energy $K_{\text{bt}}=\frac{1}{2}(\langle
v_{\theta}\rangle^2+\langle v_{\varphi}\rangle^2)$ and the baroclinic
kinetic energy
$K_{\text{bc}}=\frac{1}{2}(v_{r}^{'2}+v_{\theta}^{'2}+v_{\varphi}^{'2})$,
$\langle f \rangle$ being the radial average (barotropic component)
and $f'=f-\langle f \rangle$ the baroclinic component.

Figure~\ref{fig_ener_spec}(c) displays the power spectra for the
barotropic and baroclinic kinetic energies, as in Fig.~(3)
of~\citet{Rub_etal14}. The qualitative agreement between both figures
is remarkable. For the smaller scales the barotropic kinetic energy
$K_{\text{bt}}$ displays an $l^{-3}$ scaling as expected for a
downscale enstrophy cascade~\citep{Rub_etal14}. In contrast, at larger
scales the expected $l^{-5/3}$ scaling is replaced by $l^{-3}$ due to
the appearance of a large dipole structure. The scaling $l^{-5/3}$
observed for the baroclinic kinetic energy $K_{\text{bc}}$ is in
agreement with a three-dimensional turbulence cascade from the
convective scale. The different scalings observed for the barotropic
and baroclinic components provide an explanation for the different
scalings observed for the volume averaged total kinetic energy $K$ of
Fig.~\ref{fig_ener_spec}(a,b). At larger scales $K_{\text{bt}}$
dominates and thus $K$ exhibits the same scaling. In contrast, for
$l>30$ the baroclinic flow is dominant and thus there is an interval
for which $K$ follows the baroclinic $-5/3$ scaling.

\subsection{Large scale polar dipole regime}
\label{sec:pol_dyp}

\begin{figure*}
\hspace{-8.mm}\includegraphics[width=1.01\textwidth]{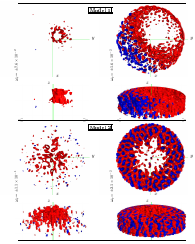}\\[-1.mm]
\caption{{\bf{Isosurfaces of radial vorticity.}}  Surfaces of constant
  radial vorticity $\omega_r$ within the tangent cylinder of radius
  $3d$ (region within the arrows in the middle line), restricted to
  the northern hemisphere. For each model two different values
  $\omega_r$ are considered, blue/red meaning $+$/$-$. On the 1st row
  of each model the surfaces are viewed from the north pole -the $z$
  axis points out of the page - to display horizontal variations. The
  point of view corresponding to an almost $\pi/2$-rotation about the
  $y$ axis of the north pole view (2nd row of plots in each model) is
  selected, to display the columnar structure.}
\label{fig_wrad_3d}
\end{figure*}

\begin{figure*}
\includegraphics[width=0.95\textwidth]{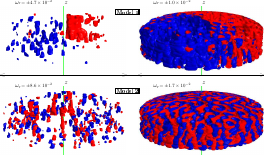}
\caption{{\bf{Isosurfaces of nonaxisymmetric radial vorticity.}}
  Surfaces of constant radial vorticity $\omega_r$ of the
  nonaxisymmetric flow within the tangent cylinder of radius $3d$
  (region within the arrows in the middle line), restricted to the
  northern hemisphere. For each model two different values $\omega_r$
  are considered, blue/red meaning $+$/$-$.  The point of view is
  selected as in Fig.~\ref{fig_wrad_3d} to display the columnar
  structure.}
\label{fig_wrad_3d_noaxi_45d}
\end{figure*}

In this section we provide numerical evidence of the formation of a
large scale dipole close to the polar regions, as
in~\citet{Rub_etal14}. In addition, we interpret the transition
between a single polar vortex (as model M1) and multiple polar
vortices (as model M2) following ideas developed
by~\citet{OEF15,OEF16} in the context of shallow layer modelling.

Coherent large scale structures are ubiquitous in the atmospheres of
gas giant planets~\citep{Fle_et_al18,Adr_et_al18}, and have been
simulated within the 2D and 3D GT context (see~\citealt{Rub_etal14} and
references therein). The latter study demonstrated the existence of a
dipole condensate in GT flows and provided an explanation of the
physical mechanism behind for its formation using a 3D reduced model
in the limit of small $\Ros$. As noted in the previous section, the
advection of 3D small scale convective baroclinic vortex produces
Taylor-Proudman (almost $z$-independent, 2D) barotropic large scale
motions, which in turn organise the small scale convective motions in
a self sustained process.

The relevant regime for the formation of the dipole condensate found
in~\citet{Rub_etal14} is $\Ray_{\text{layer}}=O(\Ek^{-4/3})$ (see
also~\citealt{Jul_etal12}), corresponding to a regime that is overall dominated
by rotation, i.~e. GT . For $\Pr=1$ a coherent
large scale dipole was found in coexistence with small scale
convective vortices. In this regime the volume renderings of axial
vorticity of~\citet{Rub_etal14} help to visualise the organization of
large columnar structures extending from the bottom to the top
boundary together with very small convective vortices. Taking into
account the differences in the definition of
$\Ray=\Ray_{\text{layer}}(1-\eta)$ the condition
$\Ray_{\text{layer}}\Ek^{4/3}=O(1)$ becomes $\Ray\Ek^{4/3}=O(0.1)$. In
our case $\Ray\Ek^{4/3}=0.023$ for M1 and $\Ray\Ek^{4/3}=0.037$ for
M2, values not so far from $O(0.1)$. As noticed in~\citet{Jul_etal12}
for small $\Pr$ the GT regime is attained for
smaller values of $\Ray_{\text{layer}}\Ek^{4/3}$ which is consistent
with our results, as $\Pr=0.01$.

The three-dimensional structure of the dipole condensate for both
models can be visualised in Fig.\ref{fig_wrad_3d} which displays the
surfaces of constant radial vorticity $\omega_r$ for two different
values (roughly half and one tenth of the maximum value). Two
different views are provided, containing the $x-y$ and the $y-z$
planes, respectively, which help us to understand the horizontal and
vertical structure of $\omega_r$. The surfaces of constant $\omega_r$
have been restricted to the volume within the coaxial cylinder of
radius $3d$ and on the northern hemisphere, for comparison with the results of the 
rotating plane geometry of~\citet{Rub_etal14}.  This
comparison makes sense as the rotating spherical shell is very thin
and thus well approximated by a plane layer (see
Fig.\ref{fig_wrad_3d}, views containing the $z$ axis).

The structure of the main polar cyclonic vortex in the case of M1 is
strongly anisotrophic, with slightly smaller horizontal scales
(compared to the vertical scale) spanning the whole shell. Small scale
cyclonic convective vortices of comparable amplitude surround the main
cyclone, some of them being very small (see the red surfaces of
Fig.\ref{fig_wrad_3d}, left column, Model 1). For small vorticity
amplitudes the anticyclonic vortices, having relatively small
horizontal scales, are less common than cyclonic vortices, evidencing a
clear asymmetry between cyclonic/anticyclonic motions. This asymmetry
was also found in the experiments of~\citet{VoEc02} investigating plane
Rayleigh-B\'enard rotating convection for $\Ros\lesssim 1$.

The surfaces of constant $\omega_r$ for M2 display similar structures
to those of M1, the main difference being the disruption of the main
cyclone of M1 into the several vortices with smaller horizontal scale
of M2, as shown previously in the contour plots of
Fig.~\ref{fig_sph_vor} in Sec.~\ref{sec:fl_pat}.  The vertical scale
of these main cyclonic vortices is of the order of the width of the
shell (see the red surfaces of Fig.\ref{fig_wrad_3d}, left column, Model
1). Again a clear asymmetry between cyclonic/anticyclonic vortices is
present in M2. The asymmetry between cyclonic and anticyclonic
vorticity is mainly due to the axisymmetric component, which is not
present for plane layer modelling~\citep{Rub_etal14} as the latitudinal
variation of the Coriolis force is not considered in their models. The
contribution to the axisymmetric component to cyclonic vorticity is
evidenced in Fig.~\ref{fig_wrad_3d_noaxi_45d} when considering only
$\omega_r$ for the nonaxisymmetric component of the flow. In this case
the number of cyclonic/anticyclonic vortices is similar, but the
cyclonic motions have larger vertical and horizontal scales.

The analysis by~\citet{OEF15,OEF16} (and also~\citealt{BSD19}) of
shallow atmospheric models has provided evidence for a transition
between different dynamical regimes which reproduces important
features seen in giant planet atmospheres. The transition between
Jupiter-like and Saturn-like polar dynamic regimes is described in
terms of the Burger number. To compare our fully convective models
with previous shallow models, the Burger number for M1 and M2 must be
estimated. For the present global simulations the spherical domain as
well as its rotation is fixed, but the characteristic horizontal
length scale $L$ of the polar convective region varies between models,
giving rise to different Burger numbers.  In our models the variation
of $L$ is achieved by modifying the Rayleigh number, i.e. the thermal
forcing, and keeping all of the other parameters (Eq.~\ref{eq:param})
fixed. In this sense the Burger number represents an output parameter
from the simulations which is computed after the flow is saturated, as
suggested in~\citet{Rea11}. This approach differs from the simulations
of~\citet{OEF15,OEF16,BSD19} in which the Burger number represents an
input parameter of the models.

According to~\citet{Ped79} the Burger number is defined as
$\Bu=\frac{g\Delta \rho D}{\rho 4\Omega^2 L^2}$, $L$ and $D$ being the
horizontal and vertical length scales and $\Delta \rho$ the density
difference between the surface and the base of the fluid. From the
Boussinesq approximation the latter can be expressed as $\Delta
\rho=-\rho_{\text{outer}}\alpha\Delta T$, and assuming $\rho\approx
\rho_{\text{outer}}$ the Burger number is
$$\Bu\approx\frac{g\alpha\Delta T D}{4\Omega^2 L^2}.$$ Considering the
full spherical shell the only quantity in this latter expression that
changes between M1 and M2 is $\Delta T$. However, as we are interested
in studying polar dynamics we should restrict the definition of $\Bu$
to the polar region. This will also permit a comparison with the local
shallow atmospheric models of~\citet{OEF15,OEF16,BSD19}. For these
reasons we take as characteristic vertical and horizontal length
scales $D=d$ and $L=\theta_{\text{polar}} r_o$, respectively,
$\theta_{\text{polar}}$ being the latitudinal angle of the extent of
polar motions on the outer surface, which clearly changes between the
models, as shown in previous sections. In terms of the nondimensional
input parameters the Burger number is then approximated as

\begin{equation}
  \Bu\approx\frac{1}{4\theta_{\text{polar}}^2}\Ray\Ta^{-1}\Pra^{-1}(1-\eta).
\label{eq:Bu}
\end{equation}

Figure~\ref{fig_wrad_3d} provides a way to infer the latitudinal
extent of polar motions in order to estimate $\Bu$ for each
model. We focus on the surfaces of constant vorticity (at
a value roughly half the maximum) viewed from the north pole that appear
in Fig.~\ref{fig_wrad_3d}. For M1, the main cyclone and its surrounding
small vortices are contained in a rectangle (on the $x-y$ plane, see
Fig.~\ref{fig_wrad_3d}) with a maximum dimension of $3d$ (in the $x$
direction). Projecting this rectangle onto the outer sphere
gives $\theta_{\text{polar}}\approx 17^{\circ}$. In contrast,
for M2, vortices having half of the maximum vorticity amplitude
spread within and over the whole coaxial cylinder of diameter
$6d$. This gives rise to a latitudinal extent
$\theta_{\text{polar}}\approx 35^{\circ}$, two times larger than for
M1. The corresponding Burger numbers (from Eq.~\ref{eq:Bu}) are
$\Bu_1=1.4\times 10^{-4}$ and $\Bu_2=5.4\times 10^{-5}$ for M1 and M2,
respectively. These values qualitatively agree with and are not so far from
those obtained in~\citet{BSD19}, which found $\Bu>10^{-3}$ and $\Bu<5\times 10^{-4}$
for Saturn and Jupiter like polar dynamics, respectively. The order of
magnitude of difference in the computation of $\Bu$ can be attributed
to the uncertainty in the determination of the Rossby deformation radius
$L_{\text{d}0}$ of Jupiter and Saturn from which
$\Bu=(L_{\text{d}0}/a)^2$ ($a$ being the osculating radius at the
poles,) was estimated in~\citet{BSD19}. According to~\citet{BSD19} the
smallest estimates of $L_{\text{d}0}$ for Jupiter may give rise to
$\Bu_{\text{J}}\lesssim 10^{-4}$, and in the case of Saturn a value of
$L_{\text{d}0}=1000$~km estimated in~\citet{CSB09} gives rise to
$\Bu_{\text{J}}\approx 2\times 10^{-4}$ (assuming $a=66810$~km). These
values are certainly in reasonable agreement with our estimations.

The physical mechanisms giving rise to a strong vortex centered at the
pole for shallow models~\citep{OEF15,OEF16} and our thermal convective
simulations share two key characteristics. Small scale baroclinic
instabilities cascade, creating a vertically aligned strong barotropic
vortex. This is demonstrated in the first part of this section as well
as in~\citet{OEF15,OEF16}. In addition, the same parameter (see
also~\citealt{BSD19}) is controlling the transition between
Jupiter-like and Saturn like polar patterns for both types of
modelling.  The authors of~\citet{OEF16} raise the possibility that
the weather layer of Saturn may be coupled with the convective
interior below. As discussed in Sec. 8 of~\citet{OEF16} convective
structures may allow local stability, which could affect the coupling
between the molecular zone below the weather layer. In this respect,
the present study is providing more support for the idea that polar
vortices are generated deep in the atmosphere, as recently discovered
for the well-known east-west low and mid latitude zonal jets
(see~\citealt{Kas_et_al18}).

\subsection{Force balances}
\label{sec:for_bal}

\begin{table}
  \begin{center}
    \begin{tabular}{lccccccccccc}
\vspace{0.1cm} Model & $\mathcal{F}_{\text{Coriolis}}$ &
$\mathcal{F}_{\text{inertial}}$ &
$\mathcal{F}_{\text{viscous}}$\\ \hline\\[-10.pt] ~~~1 & $2.0\times
10^9$ & $9.5\times 10^7$ & $8.1\times 10^5$\\ ~~~2 & $3.1\times 10^9$
& $2.8\times 10^8$ & $2.3\times 10^6$\\ \hline
    \end{tabular}
    \caption{Time and volume averaged force balance: The Coriolis
      $\mathcal{F}_{\text{Coriolis}}$, inertial
      $\mathcal{F}_{\text{inertial}}$, and viscous
      $\mathcal{F}_{\text{viscous}}$ rms forces of the two models. The
      time averages cover 1900 and 1050 planetary rotations for Model
      1 and 2, respectively.}
  \label{tab_for}
  \end{center}
\end{table}
\begin{table*}
  \begin{center}
    \begin{tabular}{lccccccccccc}
\vspace{0.1cm}           
Model & $(\mathcal{F}_{\text{C}})_r$ & $(\mathcal{F}_{\text{I}})_{r}$ & $(\mathcal{F}_{\text{V}})_{r}$ & $(\mathcal{F}_{\text{C}})_{\theta}$ & $(\mathcal{F}_{\text{I}})_{\theta}$ & $(\mathcal{F}_{\text{V}})_{\theta}$ & $(\mathcal{F}_{\text{C}})_{\varphi}$ & $(\mathcal{F}_{\text{I}})_{\varphi}$ & $(\mathcal{F}_{\text{V}})_{\varphi}$ \\
\hline\\[-10.pt]
~~~1  &  $6.1\times 10^9$          & $2.8\times 10^8$             & $6.3\times 10^6$             &  $1.0\times 10^{10}$             & $2.7\times 10^8$                 & $7.8\times 10^8$                 &  $7.3\times 10^9$                 & $1.7\times 10^9$                  & $8.0\times 10^8$ \\
~~~2  &  $1.3\times 10^{10}$       & $1.3\times 10^{9}$            & $2.3\times 10^{7}$           &  $1.2\times 10^{10}$             & $1.1\times 10^{9}$                & $2.1\times 10^{9}$               &  $2.7\times 10^9$                 & $7.2\times 10^{9}$                & $1.4\times 10^{9}$ \\  
\hline
    \end{tabular}
    \caption{Instantaneous force balance: maximum value within the
      shell of the radial, colatitudinal and azimuthal components of
      the Coriolis $\mathcal{F}_{\text{Coriolis}}$, inertial
      $\mathcal{F}_{\text{inertial}}$, and viscous
      $\mathcal{F}_{\text{viscous}}$ forces of the two models. The
      same time instant as Fig.~\ref{fig_sph_force} is considered.}
  \label{tab_inst_for}
  \end{center}
\end{table*}
\begin{figure*}
\includegraphics[width=0.95\textwidth]{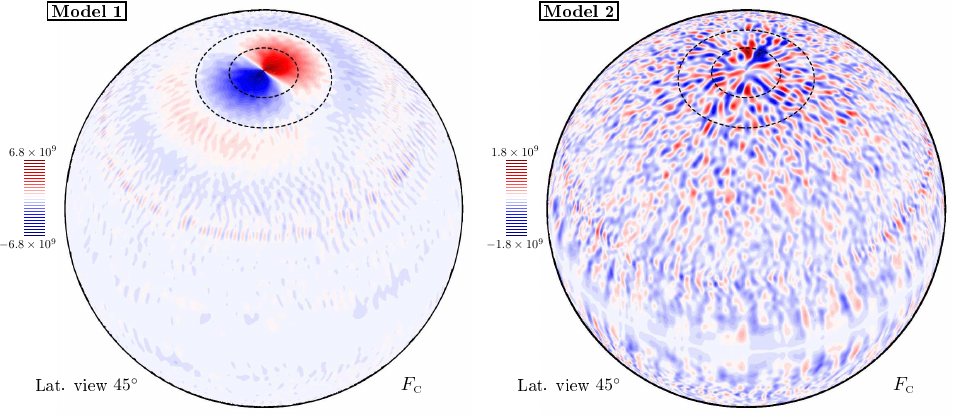}\\
\includegraphics[width=0.95\textwidth]{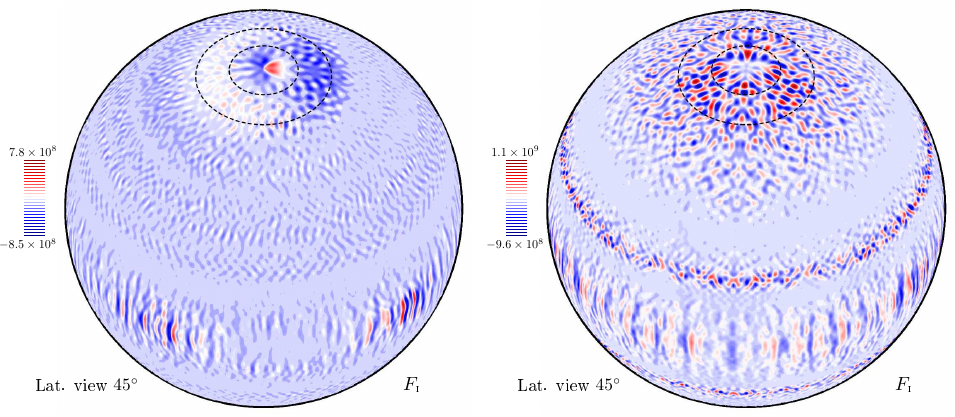}      
\caption{{\bf{Force balance of numerical models.}}  Snapshots of the
  azimuthal component of the Coriolis
  $\mathcal{F}_{\text{Coriolis}}$ (1st row) and inertial
  $\mathcal{F}_{\text{inertial}}$ (2nd row) forces, at the outer
  surface $r=r_o$ viewed from a latitude of $45^{\circ}$. The left
  column corresponds to Model 1 and the right column to Model 2. Red
  (blue) means positive (negative) values. Parallel circles at
  latitudes $-80^{\circ},-70^{\circ}$ and $70^{\circ},80^{\circ}$ are
  marked with dashed lines.}
\label{fig_sph_force}
\end{figure*}

The analysis of the force balances involved in the generation of
geophysical flows helps us to identify and characterise different flow
regimes and to shed light on the transitions occurring among
them~\citep{OrDo14b,GWA16,GOD17,JRGK12}. In the context of spherical
shell rotating convection the generation of zonal flows strongly
depends on the large scale force balance, and more specifically on the
balance between Coriolis and buoyancy forces. The nondimensional
number $\Ray^*$ was used in~\citet{AHW07} to distinguish between
rotation dominated regimes $\Ray^*\ll 1$, with strong generation of
prograde zonal flows at mid-low latitudes fed by Reynolds stresses
associated with columnar convection in spherical
geometry~\citep{Chr02}, regimes in which $\Ray^*\sim 1$ buoyancy
starts to play a role and three dimensional convection acts to
homogenise the fluid within the shell, generating retrograde zonal
flows in the equatorial region and strong prograde flows at high
latitudes. As commented in~\citet{Gla18} there are studies (for
instance~\citet{Kas_et_al18}) that have attributed the zonal flow
generation to a thermal wind balance, in which the curl of Coriolis
force balances the curl of buoyancy. In convective models of rotating
spherical shells~\citep{Gla18}, and in our models M1 and M2, this balance
does not hold as the curl of Coriolis force dominates over the other
terms.

Because $\Ray^*\ll 1$ for the numerical models M1 and M2 (see
Sec.~\ref{sec:par}), Coriolis forces are governing the global dynamics,
and prograde zonal flows are generated at low latitudes. The time and
volume averages of the forces
$\mathcal{F}=\frac{1}{V}\int_{V}(\mathcal{F}_r^2+\mathcal{F}_{\theta}^2+\mathcal{F}_{\varphi}^2)^{1/2}\text{d}V$
are shown in Table~\ref{tab_for} and verify the predominance of the
Coriolis forces. The mean values of the Coriolis
$\mathcal{F}_{\text{C}}$, inertial $\mathcal{F}_{\text{I}}$, and
viscous $\mathcal{F}_{\text{V}}$ forces verify
$\mathcal{F}_{\text{Coriolis}}>\mathcal{F}_{\text{inertial}}\gg\mathcal{F}_{\text{viscous}}$
which is the force balance believed to operate in Jupiter's convective
atmosphere~\citep{ScLi09}.

However, and in contrast to previous numerical studies
(Eg.~\citealt{AHW07,GWA13}), strong prograde zonal flows are generated
at high latitudes as well, which are preferred for $\Ray^*\sim 1$ due
to an angular momentum mixing produced by three-dimensional
convection~\citep{AHW07}. As the angular momentum depends only on the
cylindrical coordinate the zonal flow generated at polar latitudes is
still geostrophic, as in models M1 and M2. For these models the
$\Ray^*\ll 1$ condition seems to be relaxed when considering the force
balance only in the azimuthal direction, allowing the development of
high latitude prograde zonal flows. Table.~\ref{tab_inst_for}
summarises the balance for each component
$(\mathcal{F}_r,\mathcal{F}_{\theta},\mathcal{F}_{\varphi})$ of the
forces. The maximum absolute value within the shell is listed. We note
that the values are picked up at the same time instant as the
snapshots of the contour plots and surfaces of constant radial
vorticity shown previously. Whereas the Coriolis force is clearly
dominant for the radial and colatitudinal components, in the azimuthal
direction the inertial forces are comparable.

The instantaneous spatial distribution of the azimuthal component of
the Coriolis and inertial forces at the outer surface is illustrated
in Fig.~\ref{fig_sph_force} for both models at the same time instant
as the previous figures of contour plots. The figure shows that clear
correlation of both forces in the polar regions, indicating
the importance of three-dimensional
convection. Figure~\ref{fig_sph_force} also suggests that the Coriolis
predominance in the azimuthal direction is relaxed close to the equatorial
regions as well, particularly for M1. This may be the reason for the
relatively small amplitude of the prograde zonal flow of M1 and M2
(see Fig.~\ref{fig_lat_prof}) at equatorial latitudes with respect to
previous models of Jupiter and Saturn (see for
instance~\citealt{HeAu07}).

The strong dipolar character of the force balance in the case of M1 is lost for M2. The ratio
$(\mathcal{F}_{\text{I}})_{\varphi}/(\mathcal{F}_{\text{C}})_{\varphi}$
between the maximum values over the shell (see
Table.~\ref{tab_inst_for}) is $0.23$ for M1 and $2.67$ for M2. This
suggests that the value
$(\mathcal{F}_{\text{I}})_{\varphi}/(\mathcal{F}_{\text{C}})_{\varphi}\approx
1$ seems to control the transition from flows with a single polar
cyclone like M1, to flows with multiple polar cyclones like M2. At the
outer surface the situation is similar, the ratio being $0.13$ for M1
and $0.6$ for M2 (see colour palette of Fig.~\ref{fig_sph_force}).
Although the ratio for M1 is still smaller than unity, we note that in
the framework of Rayleigh-B\'enard rotating convection with parallel
horizontal boundaries~\citep{JRGK12}, which approximates our model
close to the polar regions, the GT regime is attained for values of
$\Ray^*$ which can be smaller than $O(1)$.

\subsection{Thermal transport properties}
\label{sec:th_tran}

\begin{figure}
  \includegraphics[width=0.5\textwidth]{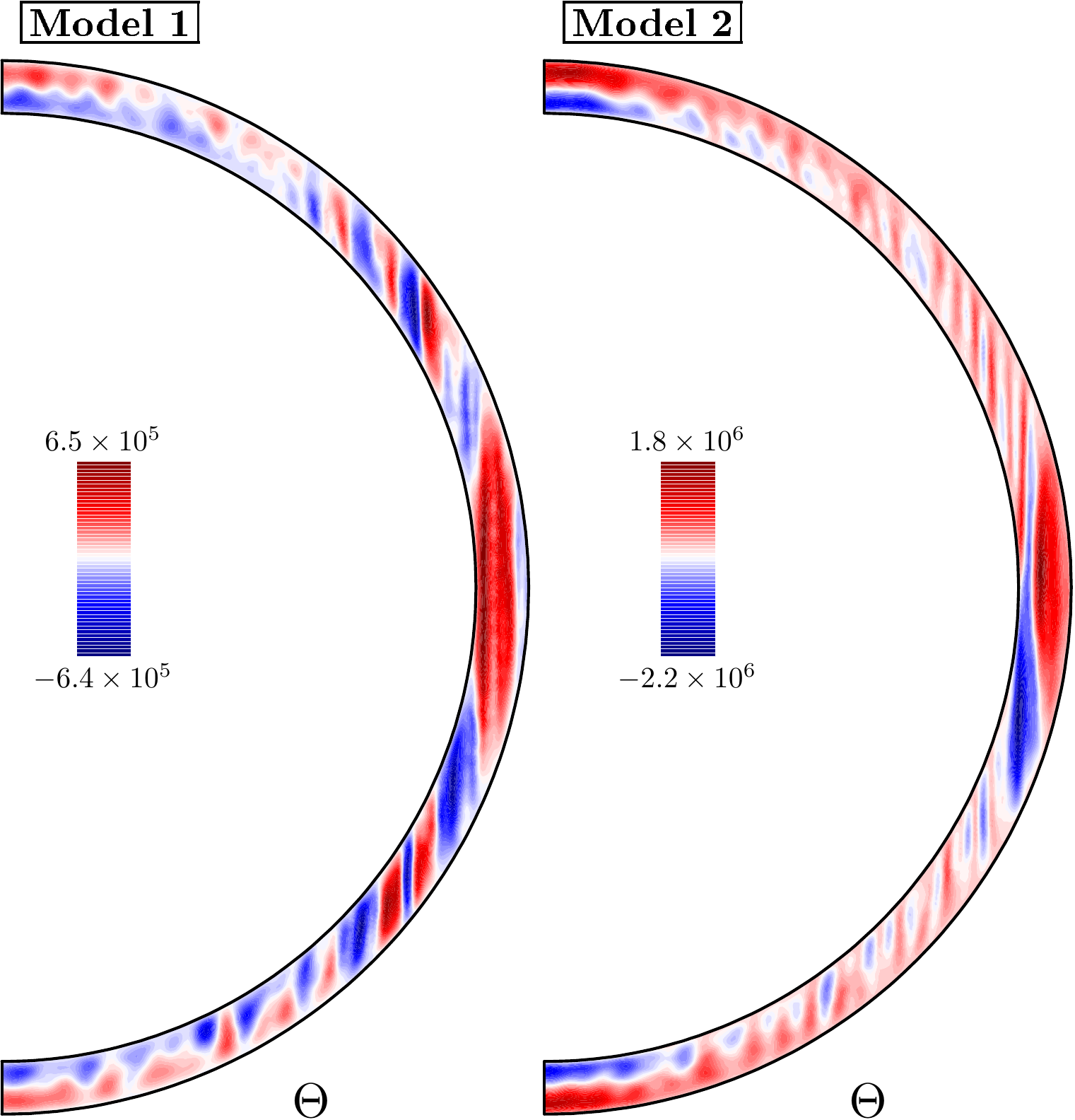}
  \caption{{\bf{Temperature meridional profiles.}}  Snapshots of the
    temperature perturbation $\Theta$ on a meridional section for
    Model 1 (left) and Model 2 (right). Positive (negative) values are
    marked with red (blue).}
\label{fig_ver_ptem}
\end{figure}

\begin{figure*}
\includegraphics[width=0.95\textwidth]{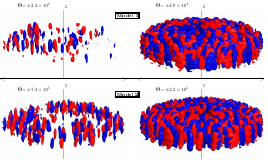}
\caption{{\bf{Isosurfaces of nonaxisymmetric temperature
      perturbation.}}  Surfaces of constant temperature perturbation
  $\Theta$ of the nonaxisymmetric flow within the tangent cylinder of
  radius $3d$ (region within the arrows in the middle line),
  restricted to the northern hemisphere. For each model two different
  values $\Theta$ are considered, blue/red meaning $+$/$-$.  The point
  of view is selected as in Fig.~\ref{fig_wrad_3d} to display the
  columnar structure.}
\label{fig_ptem_3d_noaxi_45d}
\end{figure*}

Previous studies in rapidly rotating thin spherical
shells~\citep{Aur_etal08} have analysed the heat transfer mechanisms in
the context of planetary atmospheres. In the equatorial regions the
heat flux is inhibited in the cylindrical direction as a consequence
of the strong generation of zonal flows. In contrast, in the polar
regions three-dimensional thermal plumes develop which are not
sensitive to the geostrophic zonal flow~\citep{SrJo06}, and thus at large
latitudes the heat transfer resembles that predicted for plane layer
convection. Figure~\ref{fig_ver_ptem} displays the instantaneous
contour plots of the temperature perturbation $\Theta$ on a meridional
section for both models M1 and M2. Thermal perturbations in the
equatorial regions are affected by the flow spiralling in the
azimuthal direction and extend parallel to the rotation axis in a wide
region outside the tangent cylinder. The structure of $\Theta$ close
to the poles is different, $\Theta$ being positive close to the outer
boundary and negative close to the inner, in agreement previous studies
in rotating spherical geometry of gas giant atmospheres, see for
instance Fig.~4 of~\citet{Aur_etal08}. However, in contrast to the
latter study the Nusselt number of our models is significantly
smaller, as in the polar regions the regime of GT
is already attained due to the low Prandtl number of our
models~\citep{Jul_etal12}.

For Rayleigh-B\'enard rotating convection within parallel horizontal
boundaries a recent study~\citet{Jul_etal12} provided an analysis of
the heat transfer in terms of a reduced model in the regime of
GT~\citep{JRGK12,Rub_etal14}. As described
prevously, our models M1 and M2 exhibit features, such as the dipole
condensate (Sec.~\ref{sec:pol_dyp}) and the kinetic energy spectra
(Sec.~\ref{sec:en_spec}) characteristic of GT flows,
 thus the thermal properties of M1 and M2 are in concordance
with~\citet{Jul_etal12}. Figure~\ref{fig_ptem_3d_noaxi_45d} displays
the surfaces of constant nonaxisymmetric temperature perturbation. The
latter quantity can be compared to the temperature fluctuations with
respect to the time averaged temperature (strongly axisymmetric) of
the modelling of~\citet{Jul_etal12}, yielding qualitative agreement
of the tridimensional structure of the thermal deviations (see Fig. 1
of~\citealt{Jul_etal12}).

A scaling law for the heat transport (measured by the Nusselt number
$\Nus$) was derived in~\citet{Jul_etal12} for GT flows. In this regime
turbulent 3D convection damps the heat transport in the bulk of the
fluid rather than in the boundary layers, as a consequence of the weak
$z$-dependence of the flow. The results of~\citet{Jul_etal12} point to
the independence of heat transport with respect to microscopic diffusion
coefficients, leading to the scaling
$$\Nus-1\approx C_1 \Pra^{-1/2}\Ray_{\text{layer}}^{3/2}\Ek^{2},$$
with $\Ek_{\text{layer}}=1/2\Ta^{1/2}$ for sufficiently small
$\Pra\leq 1$ and large $\Ray_{\text{layer}}$. In terms of the
definition of $\Ray=\Ray_{\text{layer}}(1-\eta)$ used in the present
study the Nusselt scaling becomes
$$\Nus-1\approx C_1
\frac{1}{4}\Pra^{-1/2}\Ray^{3/2}\Ta^{-1}(1-\eta)^{-3/2}.$$ The time
averaged Nusselt number $\Nus$, defined as the ratio of the average of
the total radial heat flux to the conductive heat flux (both through
the outer surface), has been computed for M1 and M2 in the same time
interval as the mean physical properties shown in Sec.~\ref{sec:par},
Table.~\ref{tab_param}. For M1 $\Nus_1-1=0.00827$ whereas for M2
$\Nus_2-1=0.0174$, giving rise to almost the same constant
$C_1^1=0.0296$ and $C_1^2=0.0308$, meaning that our models verify the
GT heat transfer scaling derived in~\citet{Jul_etal12}. Indeed, the
value of the constant is quite similar to the value of $C_1=0.04$ obtained
in~\citet{Jul_etal12} by means of a detailed exploration of the
parameter space, and thus our results fit reasonably well with the
theory of heat transfer developed for plane rotating Rayleigh-B\'enard
convection.

\subsection{Polygonal coherent structures}
\label{sec:pol_str}

\begin{figure}
  \includegraphics[width=0.45\textwidth]{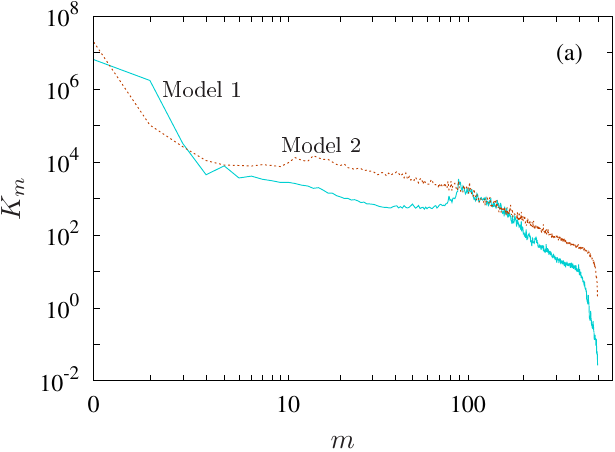}
\caption{{\bf{Energy spectra of numerical models.}}  (a) Kinetic
  energy spectra versus the spherical harmonic order $m$ for Model 1
  (solid line) and Model 2 (dashed line).}
\label{fig_enerm_spec}
\end{figure}

\begin{figure*}[h!]
\includegraphics[width=0.95\textwidth]{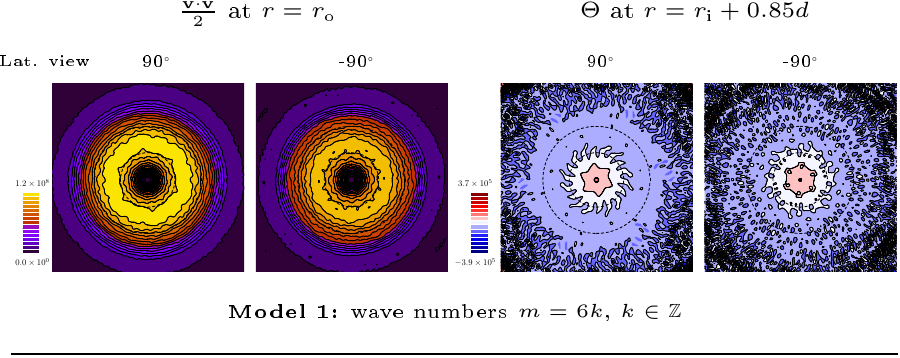}
\caption{{\bf{Hexagonal pattern for Model 1.}} Snapshots on a
  spherical surface of kinetic energy $\ve\cdot\ve/2$ (at $r=r_o$) and
  temperature perturbation $\Theta$ (at $r=r_i+0.85d$), when
  considering only the $m=6k$ ($k\in\mathbb{Z}$) component of the flow
  containing $82.7\%$ of the total rms kinetic energy. A hexagonal
  boundary extending down to $70^{\circ}$ latidude is exhibited at the
  north pole whereas this boundary is weaker and has more circular
  shape at the south pole.}
\label{fig_m6k}
\end{figure*}

\begin{figure*}[h!]
  \includegraphics[width=0.95\textwidth]{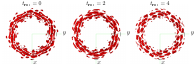}
\caption{{\bf{Isosurfaces of kinetic energy for Model 1.}}  Surfaces
  of constant kinetic energy $\ve\cdot\ve/2=7.7\times 10^7$ within the
  tangent cylinder of radius $3d$, restricted to the northern
  hemisphere. The, three snapshots with a time interval of two
  planetary rotations are shown with a north point of view. The
  projection of the outer surface $75^{\circ}$ parallel onto the $x$
  and $y$ axis is represented by a small segment.}
\label{fig_ener_3d_6m}
\end{figure*}

A significant achievement of the full-globe general circulation models
of~\citet{LiSc10} was to reproduce a meandering jet with
hexagonal-like structure, as observed in Saturn~\citep{God88}. Our
convective Model 1 generates a similar structure: belts of weak
vortices surrounding the north pole (Fig.~\ref{fig_sph_vor}, first
row, left section) appear to have polygonal shape.  The relatively
strong $m=6$ component is reflected in the plot of volume averaged
kinetic energy contained in each wave number $m$ shown in
Fig.~\ref{fig_enerm_spec}. For M1 the $m=4$ and $m=6$ components are
dominant for $m>2$ and they give rise to polygonal structures
surrounding the poles. A peak at around $m=90$ is noticeable in the
kinetic energy spectra. This corresponds to a higher harmonic multiple
of $m=6$, which is responsible for the edges and straight lines
forming the hexagonal pattern. A north pole hexagonal boundary can
clearly be identified if we consider only the component of the flow
containing the $m=6k$, $k\in\mathbb{Z}$, azimuthal wave numbers in its
spherical harmonic expansion. This flow component is strong, providing
$82.7\%$ of the kinetic energy of the flow. The contour plots of the
kinetic energy density at the outer surface ($r=r_o$) are displayed on
Fig.~\ref{fig_m6k}, the 1st and 2nd left sections corresponding to the
north and south poles, respectively. As described for Saturn (E.g. the
review~\citep{Say_et_al18}), the north hexagonal boundary is very
close to $75^{\circ}$ latitude. Our results account for a noticeable
asymmetry between north and south dynamics, which is a characteristic
feature of Saturn (see discussion in Sec. 12.4
of~\citealt{Say_et_al18}). The contour plots of the temperature
perturbation shown in the two rightmost sections of Fig.~\ref{fig_m6k}
enable us to make a qualitative comparison bewteen our results and
real measurements such as the brightness temperature maps (from
Cassini/CIRS spectroscopy) obtained in~\citet{Fle_et_al15}, with
reasonable agreement.

More than 6 years of Cassini observations have revealed that the
hexagonal pattern has rotated around 30$^{\circ}$ in the azimuthal
direction~\citep{Fle_et_al15}. Thus the pattern remains nearly
stationary in the planetary reference system~\citep{San_et_al14}. Our
preliminary investigations indicate that azimuthal drift (in the
planetary rotating frame) of the hexagonal pattern seems to be
slow. This is displayed in Figure~\ref{fig_ener_3d_6m} showing the
variation of the isosurfaces of kinetic energy density of the flow
after 2 and 4 planetary rotations, from a polar viewpoint. As observed
for Saturn (E.g. Figure 1(a) of~\citet{San_et_al14}) the hexagonal
pattern is bounding small horizontal scale vortices at around
75$^{\circ}$ latitude (marked with segments on the axis) and the
bottom vertex (that on the $x$ axis) remains stationary. The vortices
extend down to the deep interior in a spiralling fashion, so its
vertical scale is $d$.

The analysis of~\citet{Fle_et_al15} showed that the main polar vortex
and its surrounding hexagon have been persistent features on the
troposphere for over a decade, despite the seasonal evolution of the
temperature and composition. The stability of the hexagonal pattern
despite seasonal variations of insolation induced the authors
of~\citet{San_et_al14} (see also~\citealt{Fle_et_al18}) to propose its
origin as consequence of a Rossby wave extending deep into the
planetary atmosphere. The simulated deep structures for M1 persist
over more than $10^4$ planetary rotations, so the hexagonal pattern
can survive in a deep atmospheric model. This reinforces the idea of a
deep origin of the hexagonal pattern, as suggested
by~\citet{San_et_al14,Fle_et_al18}.

\section{Conclusions}
\label{sec:con}

Two fully three dimensional simulations of thermal convection in
rotating spherical shells, with aspect ratio $\eta=0.9$, are presented
in a parameter regime --defined by a Taylor number $\Ta=10^{11}$ and a
Prandtl number $\Pr=0.01$-- similar to those used previous models developed for
the understanding of moderate and low latitude wind jets observed in
giant planets~\citep{HeAu07,HAW05,HGW15}. Although the present
$\Pr=0.01$ is one order of magnitude smaller than previous models, it is
still reasonable for hydrogen-helium-water mixtures, as determined in
simulations of thermal properties along the Jupiter
adiabat~\citep{Fre_et_al12}. For the selected parameter regime the
onset of convection is in the form of nonaxisymmetric polar
modes~\citep{GSN08,GCW18} and thus polar convection is excited from
the onset~\citep{GCW19}.  This does not occur in previous modelling of
gas giant atmospheres at $\Pra\ge0.1$~\citep{HeAu07,HAW05,HGW15}, for
which the onset of convection take place at low latitudes in the form
of spiralling modes~\citep{Zha92}. When this is the case, strong
supercritical regimes are required for the development of polar
dynamics~\citep{AHW07}.

The present models, differing in the amplitude of thermal forcing
(measured by the Rayleigh number), are nonlinear, purely convective
and turbulent, but strongly geostrophic as confirmed by the time
averaged P\'eclet, Reynolds and Rossby numbers
(E.g.~\citet{KSVC17}). In addition, they exhibit a prograde zonal flow
near the equatorial region and at the outer boundary, as in gas giant
atmospheres, which is sustained by nonlinear Reynolds stresses as a
consequence of the progressive tilt of convective columns at low
latitudes~\citep{Chr02}. A new feature of these GT
flows ($\Ray^*\ll 1$) is the generation of a large amplitude prograde
zonal flow at high latitudes and the existence of strong cyclonic
vorticity surrounding the poles. Indeed, our study shows a new
transition between flows revealing strong polar activity.  This
transition differentiates two different polar regimes: at small
thermal forcing a single cyclonic vortex is found on both poles
revealing features qualitatively similar to the observed on
Saturn~\citep{Say_et_al18}, whereas at larger forcing the single
vortex is replaced by a small number of smaller cyclonic vortices
localised near the poles and thus comparable to what has recently been
found in Jupiter~\citep{Adr_et_al18}.

For the Saturn-like model the polar cyclonic vortex survives during
the whole simulation and thus seems to be a long term feature of the
flow as seen in Saturn~\citep{San_et_al14,Fle_et_al15}. By looking
carefully at the different components of the flow we observe that the
single polar vortex is bounded by an hexagonal pattern on the north,
but a polygonal boundary is absent on the south. Indeed, the hexagonal
boundary extends down to around $75^{\circ}$
latitude~\citep{Say_et_al18}. In the case of the Jupiter-like
simulation there is a small set of circular vortices surrounding the
poles, which are moving and merging. Although this contrasts with
observed structure~\citep{Adr_et_al18}, the cyclonic vortices remain
quite circular and are always present, remaining trapped near the
poles (within the latitude circles $\pm 80^{\circ}$). The long term
persistence (over two years of observation) of the circumpolar Jovian
vortices has recently been confirmed by~\citet{Tab_et_al20,
  Adr_et_al20} using Juno data. The analysis of low wave number
$|m|\le 30$ and large wave number $|m|>30$ components of the flow has
revealed that polar dynamics are governed by low wave numbers having a
multimodal structure, where large scale coherent convection is
localised in the polar regions as well as around $\pm 24^{\circ}$
latitude. In contrast, for the modes with $|m|>30$ the convective
vortices are small and only present within the equatorial band defined
by $\pm 24^{\circ}$ latitude.

An analysis of the kinetic energy spectra has allowed us to identify
the inverse cascade mechanism and to further demonstrate the GT
character of the flows. A comparison with kinetic energy spectra
obtained from velocity measurements as well as cloud morphology
obtained from Cassini observations~\citep{ChSh11} reveals a strong
similarity with our result; in particular the presence of the $-5/3$
and $-3$ scalings at low and high wave number respectively. Because
the geometry of a very thin rotating spherical shell is reasonably
well approximated by a plane rotating layer close to the poles our
results can be interpreted within the theoretical framework of GT
provided by a reduced model in the asymptotic
limit~\citep{JRGK12,Jul_etal12,Rub_etal14} of rotating plane
Rayleigh-B\'enard convection. Large scale barotropic structures are
developed as a result of the so-called spectral condensation
process~\citep{Rub_etal14}. The $-5/3$ scaling is associated with the
baroclinic convective component of the flow whereas the $-3$ scaling
comes from a downscale cascade of the barotropic component.  The large
scale polar vortices observed in our models are related to the
formation of a dipole condensate present in the regime of GT
$\Ray_{\text{layer}}E^{4/3}\sim O(1)$~\citep{Rub_etal14}. We have
analysed the three-dimensional structure of this dipole and compared
it with the results for a plane rotating layer. The large scale
dipole, which extends throughout the shell in the vertical direction,
coexists with small scale vortices of the same amplitude. Our models
favour cyclonic vorticity as observed in gas giant atmospheres. This
is also in agreement with the plane Rayleigh-B\'enard rotating
convection experiments of~\citet{VoEc02} performed for $\Ros\lesssim
1$.

The observed polar dynamics of the giant planets have been studied in
detail in the context of shallow atmospheric
models~\citep{Sco11,OEF15,OEF16,BSD19} providing a theoretical
framework, in the context of tropical cyclone theory, for
understanding their formation mechanism in giant planets (see the
comprehensive review of~\citealt{Say_et_al18} for the case of
Saturn). The main idea is that accumulation of cyclonic vorticity at
polar regions is favoured by the nonlinear advection of background
vorticity produced by the circulation inside a single vortex, the
so-called beta-gyre drift effect~\citep{Sco11}. The modelling
of~\citet{OEF15,OEF16} incorporates the effect of moist convection and
the tendency of convective storms to align in the vertical direction
and concentrate cyclonic vorticity near the poles. The mechanism
giving rise to this accumulation is strongly related to an energy
tranfer between the baroclinic and barotropic components of the
flow. In particular, small scale baroclinic vortices feed a large
scale coherent barotropic vortex~\citep{OEF15,OEF16}.  The polar
vortex formation in our modelling is produced by the same
mechanism. The transition between Saturn-like and Jupiter-like polar
dynamics has also been described by~\citet{OEF15,OEF16} and
characterised in terms of the Burger number. Our results in a rotating
spherical shell are in close agreement with~\citet{OEF15,OEF16,BSD19},
revealing rather similar dynamical behaviour. We provide an estimate
of the Burger numbers $\Bu_1=1.4\times 10^{-4}$ for the Saturn-like
model and $\Bu_2=5.4\times 10^{-5}$ for the Jupiter-like model, which
match reasonably well to the results of~\citet{BSD19}. The strong
similarities in key features between the shallow water models and the
present models may indicate a coupling between the weather (outermost
layer) and the deep convection zone beneath. This has been also
pointed out in~\citet{OEF16}.

An inspection of the relevant balances, by means of the computation of
the time and volume averaged rms forces present in our models,
confirms the global predominance of the Coriolis effect indicated by
$\Ray^*<1$ and $\Ros<1$. However, the balance in the azimuthal
direction is significantly different, and the ratio between the
inertial and Coriolis components is
$(\mathcal{F}_{\text{I}})_{\varphi}/(\mathcal{F}_{\text{C}})_{\varphi}=O(1)$. This
allows the formation of high latitude prograde zonal
flows~\citep{AHW07} which are fed from small scale tridimensional
convection thanks to an inverse cascade process. The development of
zonal circulations by means of this process was conjectured
in~\citet{JRGK12} (see conclusions) when the latitudinal dependence of
the Coriolis effects was included in their plane layer model. Our
results suggest that the ratio
$(\mathcal{F}_{\text{I}})_{\varphi}/(\mathcal{F}_{\text{C}})_{\varphi}\approx
1$ acts as a boundary separating the regime characterised by a single
polar cyclone and the regime characterised by multiple polar
cyclones. Further simulations exploring the parameter space are
required to confirm this.

The heat transport mechanisms involved in our models strongly fit with
the theory developed in~\citet{Jul_etal12}, in terms of the same
reduced model investigating the dipole condensate and the associated
energy transfer mechanisms~\citep{JRGK12,Rub_etal14}. The heat
transport within the bulk of the fluid is reduced thanks to the
efficient mixing properties of 3D small scale convection giving rise
to $$\Nus-1\approx C_1
\frac{1}{4}\Pra^{-1/2}\Ray^{3/2}\Ta^{-1}(1-\eta)^{-3/2},$$ with
$C_1=0.03$ which is equivalent to the law given in~\citet{Jul_etal12}
(there with $C_1=0.04$). Our results thus support the validity of this
scaling on the full thermal convection equations in spherical
geometry, whenever the regime of GT is attained in
the polar regions.

Summarising, the main results of this paper are the following:
\begin{itemize}
\item Thermal convection models in thin rotating shells reproduce
  qualitatively key features observed in the polar regions of Jupiter
  and Saturn:  a single polar vortex, surrounded by a hexagonal
  structure, in the case of Saturn; and an array of circumpolar
  vortices in the case of Jupiter.
\item A physical mechanism, involving a energy cascade between
  baroclinic and barotropic flows, is found to be responsible for the
  formation of polar coherent structures. This is in agreement with
  what is found in shallow weather layer models~\citep{OEF15} and
  classical Rayleigh-B\'enard geostrophic
  turbulence~\citep{Rub_etal14}.
\item The transition between single or multiple vortices is described
  in terms of the Burger number as found in previous shallow
  modelling~\citep{OEF15,OEF16,BSD19}.
\item The simulations reproduce the observed long-term stability of
  the hexagonal pattern and polar vortex as in the case of
  Saturn~\citep{San_et_al14,Fle_et_al18} and the persistence of
  circumpolar cyclones as in the case of
  Jupiter~\citep{Tab_et_al20,Adr_et_al20}.
\item The above results suggest that polar coherent structures
  observed in the weather layer of giant planets may be closely linked
  to convection in the deep interior. This has already suggested
  in~\citet{San_et_al14,OEF16,Fle_et_al18}.
\end{itemize}

Our models assume the Boussinesq approximation and thus we are not
modelling (as in~\citealt{HGW15} for instance) the stratification
occurring in real giant planetary atmospheres. However, the
incompressibility condition renders the problem more tractable, as the
numerical method does not use either any symmetry assumption or
hyperviscosity, which may facilitate the development of low wave
number flows responsible for the polar dynamics. According
to~\citet{JKM09} Boussinesq convection provides valuable information
for analysing anelastic flows. In addition, as noted in~\citet{CJM15},
when flow velocities are moderate the qualitative behaviour of
Boussinesq flows may prevail under stratified conditions, suggesting
that polar dynamics such as those exhibited by our models may be
present in anelastic models. As noted in~\citet{HAW05} (see also the
extended discussion in~\citealt{SKAI18}) the Boussinesq approximation
is still a reasonable approach for thin shells. This is further
supported in the review of~\citet{SKAI18}, where it is stated that
anelastic models provide similar flow features to the Boussinesq
approach in the case of Saturn models. The very recent study
of~\citet{CuTo20}, in a rotating plane layer, has shown that large
coherent structures may be disrupted in the upper layers with the
effect of stratification but, as concluded in~\citet{CuTo20}, it
remains unclear whether the coherence is destroyed when the global
Rossby number of the models is small, as is the case for our current
models (see Table~\ref{tab_param}).

Further research would require us to incorporate the anelastic
approximation or a nonslip condition at the inner
boundary~\citep{AuHe04}, which would be more appropriate to mimic the
damping of the zonal flow due to magnetic fields~\citep{SrJo06,LGS08},
but this is out of the scope of the present work. The presented and
previous models of thermal convection in rotating spherical shells
(Eg.~\citealt{Chr02,HAW05,HGW15}, among many others) are still far
away from the real parameters estimated for the giant planet
atmospheres. The validity of those studies relies (see the discussion
of~\citealt{SKAI18}) on the fact that they achieve a similar flow
regime to that believed to exist for the planetary atmospheres:
turbulent flows driven by large rotation (GT) and thus of small Rossby
number. In this context, the prediction of flow properties at the real
parameters is performed by means of scaling laws obtained from the
simulations (as in~\citealt{Chr02} for instance).

Further investigation is needed to track the described transition
between polar regimes in parameter space: in particular the
dependence on the Prandtl and Taylor numbers, as analysed
in~\citet{GCW18} for the onset of convection.  This is challenging, as
fully nonlinear three-dimensional spectral simulations, with large
truncation radial and angular parameters, are required to capture
polar dynamics in thin rotating shells at large $\Ta$. A comprehensive
analysis of the time scales exhibited by the polar flows will provide
valuable information in order to extrapolate to real situations and
hence improve our understanding of polar dynamics in gas giant atmospheres.


\section*{Acknowledgements}

F. G. was supported by a postdoctoral fellowship of the Alexander von
Humboldt Foundation. The authors acknowledge support from ERC Starting
Grant No. 639217 CSINEUTRONSTAR (PI Watts). This work was sponsored by
NWO Exact and Natural Sciences for the use of supercomputer facilities
with the support of SURF Cooperative, Cartesius pilot project
16320-2018. The authors wish to thank T. Gastine and J. Wicht for
useful discussions.

\section*{Data Availability}

The data underlying this article will be shared on reasonable request
to the corresponding author.

\bsp	
\label{lastpage}
\end{document}